# The Structural Biology and Critical Evaluation of Bacterial Proteases as Targets in New Drug Design

**Maria Antony Dhivyan JE,** School of Health and Life Sciences, Edinburgh Napier University, Edinburgh, United Kingdom – EH10 5DT. Email: 08800596@live.napier.ac.uk. Ph No. 07545577325.

# ABSTRACT

Bacteria produce a range of proteolytic enzymes (for which a number human equivalent or structurally similar examples exist) and the primary focus of this study was to analyse the published literature to find proteolytic enzymes, specifically endoproteses and to examine the similarity in the substrates that they act on so as to predict a suitable structural motif which can be used as the basis for preparation of useful prodrug carriers against diseases caused by specific bacteria like Salmonella. Also, the similarities between the bacterial proteases and the action of human matrix metalloproteinases (MMPs), together with the MMP-like activity of bacterial endoproteases to activate human MMPs, were also analysed. This information was used to try to identify substrates on which the MMPs and bacterial proteases act, to aid the design of oligopeptide prodrug carriers to treat cancer and its metastatic spread. MMPs are greatly involved in cancer growth and progression, a few MMPs and certain proteases share a similar type of activity in degrading the extra cellular matrix (ECM) and substrates including gelatin. Our primary targets of study were to identify the proteases and MMPs that facilitate the migration of bacteria and growth of tumour cells respectively. The study was thus a two-way approach to study the substrate specificity of both bacterial proteases and MMPs, thereby to help in characterisation of their substrates. a Various bioinformatics' tools were used in the characterisation of the proteases and substrates as well as in the identification of possible binding sites and conserved regions in a range of candidate proteins. Central to this project was the salmonella derived PgtE surface protease that has been shown recently to act upon the pro-forms of human MMP-9.

# KEYWORDS

| | |
|---|---|
| BLAST | Basic Local Alignment Search Tool. |
| BLASTp | Basic Local Alignment Search Tool for proteins. |
| DegP | Periplasmic protease |
| DNA | Deoxyribo Nucleic Acid |
| EBI | European Bioinformatics Institute |
| ECM | Extra Cellular Matrix |
| EMT | Epithelial to Mesenchymal Transformations |
| Kgp | Lysine specific cysteine protease of *P. gingivalis* |
| MMPs | Matrix Metalloproteinases |
| NCBI | National Centre for Bioinformatics |
| OMP1 | Outer Membrane Protease Of *E. clocae* |
| OMP2 | Outer Membrane Protease Of *Erwinia spp.* |
| OmpT | Outer Membrane Protease Of *E. coli* |
| PDB | Protein Data Bank |
| PgtE | Outer Membrane Protease |
| | Of Salmonella |
| Pla | Outer Membrane Protease Of *Y. pestis* |
| PrtA | Outer Membrane Protease *P. luminescens* |
| RgpA | Arginine specific cysteine Proteases of *P. gingivalis*, Chain A. |
| RgpB | Arginine specific cysteine Protease of *P. gingivalis*, Chain B. |

# 1. INTRODUCTION

## 1.1 Cancer, ECM and Proteases

Cancer has been a well-known disease since antiquity, though its relevance increased once the average life expectancy of man increased considerably due to science and technological advances, since most cancers are associated with increasing age. In a report by The American Cancer Society in 2007, cancer is thought to have caused about 13% of all human death in that year, which is equal to about 7.6 million deaths annually (Fox and Walsh, 2008). This is a compelling reason for the widespread effort throughout the world, which is being carried out to fight this menace, including preventive measures, academic research, and drug development. Generally cancer involves uncontrolled growth of cells, invasiveness and may involve the spread of cancerous cells to other parts of the body via routes such as blood; this process is called metastasis. About 150 years ago, a German microscopist suggested that solid tumors were made of a mass of cells, which was likely the starting point to study the differences between normal and tumor cells, which would eventually lead to develop treatment against cancer (Franks and Teich, 1997).

The role of non-malignant cells that are present in the close proximity of tumor cells in the progression of tumor was not well known until recently (Hanahan and Weinberg, 2000). The tumor cell's microenvironment consisting of various cells like fibroblasts, inflammatory cells, endothelial cells and myoepethial cells etc can affect the behaviour of tumor cells significantly. Apart from these cells the extracellular matrix (ECM) and its proteins create a non-cellular compartment around the tumor (DeClerck *et al.*, 2004). Different proteases produced by both non-neoplastic and neoplastic cells, as well as those present on the surface of the cells modify and remodel this non- cellular compartment, also new signals are produced and important changes in the cell to cell and cell to ECM interactions take place due to the action of these proteases (Liotta and Kohn, 2001). Important cellular characteristics such as survival, motility, differentiation, and proliferation are affected ultimately by these signals via altering the gene expression in the cell (Werb, 1997).

EMT or Epithelial to Mesenchymal Transformation is the process in development by which the basolateral polarity and strong intracellular adhesion features of epithelial cells are lost, and by which in turn they gain ability to move through the ECM (Savagner, 2001). As cells lose contact with each other the interactions between cells and the ECM are generated, which leads to change in function of several adhesion molecules present on the cell surface, which might now become the source of various signals promoting motility, growth, ECM degradation and metastasis.

The role of proteases in cancer has been substantially studied, where the main focus of various research groups has been to assess the critical role of proteases in the degradation of the ECM, which facilitates the penetration of tumor cells into blood vessels and surrounding connective tissue (Koblinski *et al.*, 2000). In particular the effect of a certain class of proteases called MMPs or matrix metalloproteinases has been widely studied and their role in ECM degradation analyzed. Although MMPs are essential in the breakdown of ECM in normal physiological processes such as tissue remodelling, reproduction and embryonic development, their role in tumour progression is more widely studied (Nagase and Woessener, 1999). For example, MMP-7 or matrilysin is widely expressed by breast tumour cells and its been observed that it causes visible breaks in the basement membrane by cleaving matrix components in the cellular environment (Noe et al., 2001). However MMP- 7 is not only involved in the breakdown of ECM in cancer but also has other functions; it also cleaves E- cadherin, which is a cell adhesion molecule, thereby causing epithelial cell disruption, increased invasiveness of the cell and high saturation density (Fingleton et al., 2001). Likewise all the MMPs have different functions relating to tumour development along with the degradation of the ECM (Fridman, 2006).

Proteases are normally expressed as inactive forms in the extracellular environment and are activated by various mechanisms including peptide cleavage. However many of the antibodies are unable to distinguish between the active and inactive forms of proteases, so immunohistometry studies showing over expression of proteases doesn't always means there is increased proteolytic activity (Coussens et al., 2002). These proteases have been traditionally considered as targets for anti- cancer treatments because of their obvious involvement in tumour growth and progression. Most studies on targeting proteases for cancer treatments involve the following standpoints, finding the proteases that are involved in malignant progression, validation of the found proteases as possible targets for therapy, development of assays that help to analyze the proteolytic activities and cellular interactions inside a solid tumor and also to develop non invasive imaging techniques to study the actions of these proteases and thereby developing suitable anti-protease therapies (Mahmood and Weissleder, 2003). There are many natural protease inhibitors in nature, which effectively inhibit the function of proteases; alpha-1 antitrypsin is a well-known protease inhibitor whose deficiency is associated with emphysema and liver disease (Crystal, 1990). However there has also been a problem with synthetic inhibitors of proteases for cancer treatment that have not been deemed efficacious in clinical trials; for example, preclinical studies of MMPIs or the MMP inhibitors have shown that they are efficacious in prevention techniques or during early stages of tumor growth but not in later stages of cancer. MMP inhibitors have been used like cytotoxic drugs, but actually they are cytostatic in nature, so they won't cause tumour shrinkage at later stages of cancer, though they maybe useful as preventive measures (Egebald and Werb, 2002). Also an efficient method for systemic delivery is not available, which poses a further problem for utilising this technique (Brown, 1997). Further the use of protease inhibitors have also known to have some side effects, which have been observed in treatment against cancer, bacterial diseases and AIDS (Tsiodras *et al*., 2009).

## 1.2 Matrix Metallopeptidases

The Matrix Metallopeptidases or MMPs as they are commonly known are a class of endopeptidases containing catalytic Zinc group and their primary function involves the degradation of the ECM and its components. A common property of the MMPs include that they are generally in the form of inactive zymogens which have to be activated by proteolytic cleavage (Harper et al., 1971). Normally they are expressed only during processes such as wound healing, reproductive functions like embryonic development, placenta development etc. However, abnormal functions of MMPs have been connected to various diseases such as rheumatoid arthritis, gastroenteritis, periodontitis and cancer (Woessener et al., 1991).

The structure of MMPs normally is made up of a catalytic domain, which is about 180 amino acids long, a propeptide region of about 80 amino acids, a linker or hinge region of variable length and a hemopexin domain equal to about 200 amino acids (Jones *et al*., 2003). However MMP-7, MMP-23 and MMP-26 are exceptions to the general structural formula since they don't have the hemopexin domain and hinge region. MMP-23 also possesses a unique region rich in cysteine and an immunoglobulin domain, further, MMP-2 and MMP-9 have three repeats of a fibronectin motif in the metalloproteinase domain (Nagase et al., 2006).. There are some common structural signatures are there as well among MMPs, in the catalytic domain there is a zinc binding motif HEXXHXXGXXH and also there is the motif PRCGXPD known as cysteine switch in the propeptide region. The cysteine and zinc coordination is required to keep the MMPs in an inactive state; a Met-turn is also formed with a conserved methionine present in the catalytic domain, which provides the support for the catalytic zinc in the form of a structural base (Bode et al., 1993).

### 1.2.1 MMPs and Bacterial Diseases

MMPs have proteolytic affinities that are found to be overlapping and they can cause degradation of the extracellular matrix. MMPs are also divided into elastases, collagenases, stromelysins and gelatinases based on their substrate specificity. These substrates also contain cytokines, proteinases, chemokines and inhibitors of the proteinases along with the ECM components (McCawley and Matrisian, 2001). The Cysteine and Zinc binding maintains the inactive form of MMP and an intermediate activation could be caused by even other MMPs or by molecules such as Trypsin and Plasmin. But for complete activation another cleavage has to happen, which might be autocatalytic (Johansson et al., 2000). Many MMPs' expression is brought about by various outside signals such as contact with surface components of microbes or the ECM (Opdenakker et al., 2001). It has been observed that host cell like monocytes produce MMPs during invasion by viruses and bacteria (Elkington et al, 2005). Previous studies have shown that some bacteria activate host proMMPs through their proteolytic enzymes, which includes bacteria such as *Vibrio cholerae*, *Treponema denticola*, *Pseudomonas aeruginosa*, *pneumoniae* bacteria (Okamoto et al, 1997) and Porpyromonas gingivalis (Pattamapun et al, 2003). The LPS from enteric bacteria such as Salmonella also induces macrophages to secrete proMMP-9 (Xie et al., 1994). Also gelatinase and collagenase production has been observed in some bacteria (Watanabe, 2004). It has also been observed that mice deficient in MMP- 3 have comparatively high resistance against systemic *Salmonella* infection than normal mice (Handley and Miller, 2007) and MMP-9 deficient mice have high resistance towards enterocolitis caused by *Salmonella* (Castaneda et al, 2005). All these results have shown that there is a probable case of MMPs taking part in *Salmonella* infections.

### 1.2.2 MMPs and Cancer

Angiogenesis is an important part of cancer progression, it involves various steps including degradation of the basement membrane on an existing blood vessel, migration and proliferation of endothelial cells into the new space, maturation, differentiation, and adherence of the endothelial cells to each other, and lumen formation. The initiation of angiogenesis may be brought about by various factors from cells such as macrophages, mast cells, inflammatory cells or tumor cells (Brooks, 1996). They bind to the matching cell surface receptors and among other things such as activation, proliferation, migration and invasion functions of the cell; they also induce the secretion of MMPs (Eliceiri and Cheresh, 2001). MMPs are involved in angiogenesis, in processes such as degradation of the ECM, its components and the basement membrane, so that the endothelial cells can migrate to new tissue. This process also causes integrin intracellular signalling, for example, it has been observed that MMP-2 initiates integrin signalling and may help in endothelial cell proliferation and survival. MMPs also generate other endogenous angiogenesis factors, such as angiostatin, endostatin and in the case of MMP-2 even the generation of a hemopexin by autoactivation (Stevenson, 1999). This shows that MMPs have both pro and anti angiogenic activities, however generally MMPs help in angiogenesis and MMP inhibitors suppress angiogenesis in experimental models (Naglich et al., 2001). During cancer progression in humans, especially in advanced stages, during invasive progression, metastasis and poor prognosis, MMP expression has been found to be increased (Coussens et al., 2002)). Early MMP expression, which might be caused by neighbouring cells or the tumor cells themselves, might help in ECM degradation and setting up of the microenvironment for cancer growth and primary establishment. They remove adhesion sites, expose new binding sites, break cell-matrix or cell-cell receptors and also release chemoatttractants from the extracellular matrix thereby aiding in the migration of tumor cells (McCawley and Matrisian, 2001). Along with intravasation MMPs are also required for extravasation of tumor cells from the blood cells, although this process does not seem to have a rate limiting effect in metastasis (Chambers and Matrisian, 1997). This shows that in the carcinogenic process MMPs are involved in multiple different stages. There is also a switch that accompanies the tumor growth at a certain point whereby the expression of pro angiogenic factors start outweighing the expression of angiogenic inhibitors; different model systems have implicated both MMP-2 and MMP-9 in causing this angiogenic switch (Egeblad and Werb, 2002). The idea for developing MMP inhibitors for cancer treatment has been around for a considerable time, though good results have been obtained from animal studies but converting these results into clinical trials in humans has not been as promising. The reasons for this can be associated with the poor bioavailability MMP inhibitors of the first generation and second-generation MMP inhibitors causing unwanted side effects (Overall and Otin, 2002). From the known roles of MMPs in the entire carcinogenesis process it could be hypothesized that MMP inhibitors might have betted effect if administered to patients at early stages of cancer development. Though study on inhibiting

MMPs to treat cancer has been widely studied, there has been comparatively a meagre amount of study on utilizing these proteases for drug delivery and chemotherapy purposes.

### 1.2.3 MMPs in Drug Delivery

The primary focus of drug delivery for many has been to find a ‘’magic bullet’’ (Ehrlich and Bolduan, 1906). If we are able to devise delivery systems that would allow drugs to target the necessary site leaving out other parts of the body, that would help to destroy that particular disease. Unfortunately, this has not been the case most diseased tissues remain to normal except that they show increased or decreased level of the biological processes compared to normal tissues. This gives rise to the idea to find the biological processes or targets in which most difference is seen between normal and diseased tissues, so as to differentiate them, one such target is the MMP. Tumours are surrounded by a unique microenvironment, an over production of proteases (Egeblad and Werb, 2002) and slightly increased acidic pH could be observed generally (Gerweck and Seetharaman, 1996). And MMPs are overproduced as well, since they are essential in the carcinogenic process. Previously, to locate or identify tumours as well as to turn on or activate imaging agents, this over expression of MMPs has been utilized (Weissleder et al., 1999; Mok et al., 2009). The same has been done with imaging based therapy for other diseases such as arthiritis (Lee et al., 2008) and monitoring response for treatments (Wunder et al., 2004). On of the very recent methods of drug delivery has been the use of custom created nanocarriers that transfer a drug payload to the target site. And the optimum way to do that is to keep them inactive until they reach the target site for efficient target delivery. The activation of these nanocarriers could be programmed, so that, they get activated by factors in the tumor microenvironment, to increase successful targeting. The acidic pH as well as the over expressed enzymes could be utilized for this purpose. By using a peptide linker to connect methotrexate and dextran a drug polymer conjugate was formed, which was observed to be cleaved by MMP-9 and MMP-2 (Chau et al., 2004). But in this study it was observed that there was not much difference in the uptake of drugs from MMP sensitive and MMP insensitive drug polymer conjugate. Though the reason for this might because the cleavage of the linker happened there was endocytic uptake of the conjugate itself (Chau et al., 2006). Also the efficacy of MMP sensitive and insensitive conjugates was much higher compared to free drug in various MMP over expressing tumor models (Chau et al., 2006). Another study reports the idea to use a MMP sensitive peptide to link 1,2-dioleoyl-sn-glycero-3-phosphoethanolamine and polyethylene glycol (PEG), this study showed in MMP over expressing tumors MMP sensitive conjugates induced three times higher gene transfection than done by MMP insensitive conjugates (Hatekeyama et al., 2007). The structure of the various MMPs differ significantly which might enhance their substrate specifity, the catalytic domains contain various residues and is responsible for catalysis and binding, though other discrete binding domains exist as well in other regions in the MMP, which further increases specificity for substrates (Overall, 2002). To find the substrate specifity of MMPs various methods such as genetic knockout, proteomics based methods and affinity methods could be used (Tam et al., 2004). If we consider the cleavage site of a substrate for MMP as

…P3 - P2 - P1 ~ P1' - P2' -P3'… where P could represent any amino acid and cleavage will happen between P1 and P1'. MMPs generally prefer small residues such as Serine at P1, hydrophobic residues at P1' and P2, basic or hydrophobic residues at P2' and Proline at P3. Though most of these preferences are common for most MMPs, there has also been some differences, for example, MMP-1, MMP-2 and MMP-9 prefer small residues at P3', whereas MMP-3, MMP-7 and MMP-14 prefer methionine (Vartak and Gemeinhart (2007), This way by analyzing substrate specificities of the MMPs its possible to design substrates that will be specifically cleaved by selected MMPs and could be used for targeted drug delivery (Turk et al., 2001).

## 1.3 Bacterial Endopeptidases

Endopeptidases can be classified as a class of peptide hydrolases that cleave peptides or proteins internally, the peptide bonds of non-terminal amino acids are broken in contrast to other peptidases that break bonds from the terminal pieces (Cassarini *et al*., 1999). Endopeptidases cannot cleave peptides to form monomers and they are normally identified based on the specifity in which they cleave. For example, Trypsin is a very specific endopeptidase, which always cleaves after Lysine or Arginine, unless when either is followed by Proline (Leiros *et al*., 2004). Endopeptidases are further classified in to serine proteases, threonine proteases, cysteine proteases, aspartic acid proteases and metalloendopeptidases based on the amino acid present in their active site. In the case of metalloendopeptidases the active site includes a metal such as zinc and at times cobalt and manganese (Hashimoto, 1983). The aspartic acid proteases and metalloproteases are different from the other proteases in that their nucleophile in the catalytic site consists of a water molecule compared to the amino acid of the other proteases (Barrett et al., 1998).

Bacterial produce a variety of proteases and their role in virulence factors of the various bacteria are very evident. Its well known that bacterial proteases have the ability to destroy important host defence proteins, such as those present in the complement system which is the main part of a host's innate immunity (Popadiak et al., 2007), as well as degrade the functional and structural proteins present in the host organism (Maeda, 1996). More specific roles for bacterial proteases have recently been observed in the interaction between the host organism and the invading bacteria that includes, blocking of cascade activation pathways and cytokine network, cleavage of receptors found on the surface of the cell, and inactivation of the protease inhibitors present in the host organism (Rice et al., 1999). Also it should be noted that the ability of bacteria to use the proteases it produces to pass the proteinaceous barriers present inside a host organism is a major part of bacterial virulence. A more interesting fact is that bacteria use proteases to acquire or activate the function of host proteases to help in its growth or progression of disease (Lantz, 1997). Bacterial protease can be targeted for producing new type of antibiotics, since the inhibition of their functions would result in the death of the invading bacteria (Travis and Potempa, 2000). Though targeting proteases for creating inhibitors has proven to be a useful idea for therapeutic intervention against bacteria, utilizing these proteases or taking advantage of their functions to help in drug delivery techniques have not been widely tried.

### 1.3.1 *Salmonella* and PgtE

*Salmonella* spp. based diseases cause more than 3 million deaths annually and it is one of the most important food borne infections all over the world (Pang et al., 1995). *Salmonella* consists of two species with very different characteristics regarding effects on human beings. *S. enterica* is highly pathogenic to humans as well as most warm-blooded animals, unlike *S. bongori*, which rarely causes human diseases (Giammanco *et al*., 2002). *Salmonella* and Escherichia coli are genetically related and they share approximately 80% similarity in their genomic sequence (McClelland et al, 2001). Also has been hypothesized that *Salmonella* and *E. coli* may have diverged from an ancestor some 120 million years before, after which they developed various phenotypic changes that differentiate these from each other (Ochman and Wilson, 1987). *Salmonella enterica* serovars Typhi causes the dreaded typhoid, which alone causes over 200000 deaths a year. *Salmonella* enters the body through the ingestion of contaminated food or water and it goes through a series of invasion processes including crossing the intestinal epithelium and infecting dendritic cells and submucosal macropages. Macrophages act as important phagocytes and effectors of the adaptive and innate immune responses. *S.* Typhimurium can survive within macrophages intracellularly, where it undergoes the process of replication within certain specialised vacuoles (Dalfors et al., 1997). During the infection process, it causes the death of the host cells and thus spreads to new cells, evidence also suggests that Salmonella disseminates extracellularly, for which it should have proteolytic activity (Knodler and Finlay, 2001).

Salmonella enterica possesses an outer membrane protein called PgtE; it is a member of the omptin family of proteases (Ramu *et al*., 2008). We know about the functions of certain omptins such as Pla, an outer membrane protein of Yersinia pestis, which plays an important role in the invasive character of plague (Kukkonen and Korhonen, 2004). Furthermore Pla has nearly 75% sequence similarity with PgtE in predicted amino acid sequences according to previous studies (Sodeinde and Goguen, 1989). OmpT of *E. coli* is another omptin that has been widely studied and whose structure was obtained through crystallography techniques; it also shares a considerable sequence similarity with PgtE (Vandeputte- Rutten et al, 2001). Regarding the proteolytic activity of omptins they are observed to cut after the basic residues of the polypeptides from the case of OmpT (Dekker et al, 2001).

PgtE degrades alpha-helical CAMPs, which shows that it might play a role in resistance towards innate immunity of *Salmonella* (Guina et al., 2000). Also PgtE was identified in degrading viral epitopes, which were fused within a Typhimurium strain, in its fimbriae and subsequent increase in antibody response was observed by removing PgtE (Chen and Schifferli, 2003). Such results show the different types of substrates that PgtE is capable of cleaving. PgtE has been found to convert plasminogen to plasmin and also is involved in the inactivation of plasmin inhibitors (Lähteenmäki *et al*., 2005). This function is similar to that of Pla, which also inactivates plasmin inhibitors and is more proficient in the conversion of plasminogen (Kukkonen et al., 2001).

Recent studies have shown the interesting property of PgtE to be involved in gelatin degradation and proMMP-9 activation. The same study tried to analyze whether Pla of *Y.pestis* and OmpT of *E.coli* had a similar effect, but showed that Pla degraded gelatin poorly and there was no proMMP-9 activation. Similarly OmpT was also poor in gelatin degradation and proMMP-9 activation. These results suggest that *Salmonella* might have separate substrate specificity to gelatin and proMMP-9 (Ramu et al., 2007).

By analysing the unique substrates on which PgtE acts on, as well as the study of differences in the sequences of PgtE and omptins like Pla and OmpT, might be possible to predict a general structural motif for a substrate on which PgtE might act. Further homology modelling of PgtE using the structural data of OmpT might also prove valuable to study the important active sites, loops, which are involved in substrate specificity. Converting OmpT into a Pla-like enzyme has been done by substituting critical amino acids that are exposed on the surface of the bacteria, this shows that the loop structures on the surface play an important role in the specificity of proteolytic properties in omptins (Kukkonen et al, 2001).

### 1.3.2 Photorhabdus and PrtA

*Photorhabdus* is the only genus of terrestrial bioluminescent bacteria that has been identified and is divided into three known species, *P. luminescens*, *P. temperata* and *P. asymbiotica* (Le Saux et al., 1999). Photorhabdus produces a certain type of proteases of the class serralysins, which are thought to have pathogenecity functions and help in the infection process of the bacteria though there is no clear knowledge of serralysin substrates (Marokhazi *et al.*, 2006). Serralysins are extracellular metalloendopeptidases that show complete conservation of $Zn^{+2}$ binding domain, Met-turn and $Ca^{+2}$ binding repeats (Baumann 1998). PrtA is one such serralysin, which is secreted by the bacteria *Photorhabdus luminescens*, which is found to be extremely pathogenic in insects. PrtA is produced during the start of the infection, which might because of its role in pathogenecity (Bowen et al., 2000). Also it is to be noted that PrtA is not involved in the bioconversion of host tissues, which would rule out its specificity (Daborn et al., 2001), neither does it have identifiable direct toxicity unlike several other mettaloproteases (Bowen et al., 2003).

To study the proteolytic system of PrtA and understand its function in the infection of *Photorhabdus*, the knowledge of its substrate is required. To study the cleavage site specificity of PrtA, the sequence was analyzed with some biological peptides such as insulin A, insulin B and beta-lipotropin. A preference for hydrophobic amino acids at positions P2, P1’ and P2’ was observed on cleavage site analysis of the substrate. From the results, it was also concluded that the cleavage sequences PrtA are rich in Ala, Leu and Val (Markokhazi et al., 2007). A simple probability analysis of amino acid frequencies indicated a slightly higher frequency of Leu and Val at position P2’, which is in accordance with the presence of a conserved Leu (Leu3, a position equivalent to P2’) of the known bacterial inhibitors of serralysin-like proteases (Bae et al., 1998). Searching for target proteins using hemolymph of *M. sexta* in the proteolytic system of PrtA has given some results that would be useful, totally 16 PAT proteins were found, which were cleaved readily and selectively by PrtA. Of the sixteen, 6 were known proteins that have some immune related activity. From this it could be supposed that PrtA participates in immunosuppressive mechanisms of *Photorhabdus*, though there is little information about the other 10 PAT proteins (Felfoldi et al., 2009).

1.2.3 *P. gingivalis* and the Gingipains.

The bacteria *Porpyromonas gingivalis* is considered to be the important etiological agent for the infection and development of chronic periodontitis, which is a severe inflammatory disease, which affects the tissues around the teeth in adults (Fox, 1992). The pathogenecity of the bacteria is associated with various factors such as, the hemagglutinins, fimbriae, LPS and some cysteine proteinases called Gingipains and their adhesive factors (Simpson et al., 2000). In these factors involved in *P. gingivalis* pathogenecity, the Gingipains, that is, Arginine specific cysteine proteinases (RgpA and RgpB) and Lysine specific proteinases (Kgp) are considered to be really important since they degrade a number of host proteins and disrupt the hosts defensive mechanism (Travis et al., 1997). Of the several functions the Gingipains possess, they are able to induce the clotting of blood by causing cascade of coagulation at different levels (Imamura et al., 2001), and the thrombin that is produced during the clotting process is known to be a proinflammatory mediator (DeMichelle et al., 1990). These two functions are responsible for the inflammotary effects associated with periodontis. Further Gingipains can degrade fibrinogen (Imamura et al., 1995) and induce the production of collagenase through gingival fibroblast resulting in activation of proMMPs (DeCarlo et al., 1997). Both Rgp and Kgp have been observed to degrade type I collagen directly in mutant studies (Tokuda et al., 1998) and Rgp can induce fibroblasts to produce MMP-1 and Kgp can induce the production of MMP- 8 (Fitzpatrick *et al*., 2009).

# AIM

The primary aim of this study is to research published literature on the proteases of interest, analyse their substrate specificities and recommend the requirements for a possible structural motif in the development of carriers for prodrugs. Basic bioinformatics tools will be used to analyse the sequential and structural data of these proteases. The results obtained will be correlated with the data obtained from the literature search to arrive at a consensus for the recommendation in the design of a structural motif for a possible prodrug carrier. This study includes the following steps,

- Extraction of the protein sequences for the proteases under study from the GenBank repository.
- Alignment search to find similar sequences using BLASTp.
- Research published literature for related information on the selected proteases having sequence similarity.
- Multiple sequence alignment between the various proteases using ClustalW.
- Search and identification of important motifs present in the primary protease under study.
- Homology modelling of the structure for the proteases under study using Swiss Model server.
- Structure Alignment and Analysis.
- An organised report on the knowledge obtained from this study towards creating a structural motif for the purpose of creating prodrug carriers using the studied proteases.

## 2. MATERIALS AND METHODS

2.1 Bacterial Strains and Proteases.

The strains and proteases of the primary bacteria studied in this work are described in Table 1. It also includes the other bacterial strains and proteases used by various studies used as the source of primary literature in this study.

| PROTEASE (Accession No) | SOURCE ORGANISM | REFERENCE |
|---|---|---|
| PgtE (AAF85951.1) | Salmonella enterica serovar Typhimurium | Guina et al., 2000 |
| Pla (NP_995574.1) | Yersinia pestis | Revazishvilli et al., 2008 |
| OmpT(NP_287408.1) | Escherichia coli | Skyberg et al., 2006 |
| PrtA (NP_928000.1) | Photorhabdus luminescens subsp. Laumondii TT01 | Duchaud et al., 2003 |
| RgpA(YP_001930085.1) | Porphyromonas gingivalis ATCC 33277 | Naito et al., 2008 |
| RgpB(YP_001929582.1) | Porphyromonas gingivalis ATCC 33277 | Naito et al., 2008 |
| Kgp(YP_001929844.1) | Porphyromonas gingivalis ATCC 33277 | Naito et al., 2008 |

### 2.2 Sequence Alignment.

This process involves the method of arranging DNA, Protein or RNA sequences to find similar regions among the sequences, which may indicate evolutionary, structural or functional relationships among them (Mount, 2004). Sequence alignment techniques are widely used in all forms of genomics and proteomics studies. For example a similar study based on finding structural and functional relationships among bacterial protease precursors used an extensive amount of sequence alignment techniques for analysing the different bacterial protein strains under study (Serkina et al., 2001).

#### 2.2.1 BLAST

BLAST identifies the local regions of similarity between different sequences, it can be used to compare protein or nucleotide sequences to the sequences already present in databases to identify similarities and provide a statistical idea about the evolutionary and functional relationships between them (Altschul *et al*., 1990). The protein sequence of the primary bacterial protease under study, the surface protease PgtE of *Salmonella enterica* serovars was obtained from the Genbank database (accession no. AAF85951.1). It is an outer membrane protease, 312 amino acids long and is taken from the *Salmonella enterica* subspecies, *Salmonella enterica* serovars Typhimurium (Guina et al., 2000). A local alignment search was used on the obtained sequence using BLASTp to compare it with sequences in the GenBank repository. Both these services were accessed through the NCBI website (http://www.ncbi.nlm.nih.gov/). The BLAST results were analyzed and the highly similar sequences were isolated so as to study the published literature to find if they possess any functional similarity and structural similarity to the protease PgtE of *Salmonella enterica* serovars.

The protein sequence of the second protease under study the alkaline metalloprotease PrtA of *Photorhabdus luminescens* was obtained from the GenBank database (accession no. CAE12950.1). Its an alkaline metalloprotease similar to serralysins, 480 amino acids long and has been isolated from the bacteria *Photorhabdus luminescens* subspecies laumondii, strain TTO1 (Duchaud et al., 2003). A BLASTp search was run using the NCBI server (http://www.ncbi.nlm.nih.gov/) and the similar sequences were analysed to find functional and structural correlation to the protease under study using published literature.

The same procedure was repeated with the gingipains, RgpA, RgpB and Kgp, the protein sequences for which were obtained from the Genbank repository. The arginine specific cysteine proteases, RgpA (accession no. YP_001930085.1) and RgpB (accession no. YP_001929582.1) were isolated from the bacteria Porphyromonas gingivalis strain ATCC 33277, and were 1703 and 736 amino acids long respectively. The lysine specific cysteine protease Kgp (accession no. YP_001929844.1) was also obtained from the same strain of the bacteria and was 1723 amino acids long (Naito et al., 2008). BLASTp was again used to study the similar sequences of these gingipains and identify any sequences that might have functional similarity using an extensive search of published literature.

The NCBI server (http://www.ncbi.nlm.nih.gov/) was mainly used throughout the BLAST analysis process.

### 2.2.2 CLUSTALW

Multiple sequence alignment is used to align two or more sequences (Wang and Jiang, 1994). It can be used to find conserved regions among the sequences of study and this can be used to identify similar structural motifs among different sequences thereby providing information to isolate active sites and other important regions. Multiple sequence alignment can be further used to create phylogenetic trees and gauge the evolutionary distances between various sequences; this proves very helpful to study the evolutionary relationships between related sequences (Chenna et al., 2003). ClustalW is one such multiple sequence alignment tool, which is used to align DNA or protein sequences and is very useful in identifying similarities, differences and identities among diverse sequences (Thompson et al., 1994). Phylograms and cladograms can also be generated to study the evolutionary relationships between sequences. ClustalW can be accessed using the European Bioinformatics Institute web server (http://www.ebi.ac.uk/Tools/clustalw2). It's a widely used tool and is used in a variety of bacterial protease study as well like the study on the effects of overproduction or absence of periplasmic protease DegP in bacterial growth (Onder *et al*., 2008).

The PgtE protein sequence obtained was aligned with two other omptins known to have sequence similarity, the surface protein Pla of Yersinia pestis, which is a major virulence factor of plague and outer membrane protease

OmpT of *E. coli* using ClustalW2. The Pla protein sequence was obtained from the Genbank database (accession no. ABX84823.1), isolated from Yersinia pestis Angola strain and measuring 312 amino acids long (Eppinger et al., 2010). The OmpT sequence was obtained from the Genbank repository (accession no. ABA54754.1), isolated from Escherichia coli and measuring 317 amino acids long (Skyberg et al., 2006). The results obtained were analyzed to identify conserved regions among the protein sequences and significant differences as well.

To analyze the substrate specificity of surface protease PrtA of *Photorhabdus luminescens*, the 6 known proteins of the actual 16 it was observed to cleave, were subjected to multiple sequence alignment, so as to identify any similarity in the sequential pattern of the substrates. This would help to hypothesize a suitable structural motif on the substrate for PrtA cleavage. ClustalW2 web tool was used to create the multiple sequence alignment through the EBI web server (http://www.ebi.ac.uk/Tools/clustalw2). The results obtained were analyzed to find any conserved regions among the sequences. The Phylograms and cladograms constructed were also accessed to ascertain the evolutionary significance.

## 2.3 Motif Identification

A sequence motif is the highly conserved region in an amino acid or nucleic acid sequence that is widespread and can be hypothesized to have some biological significance (Stormo, 2000). Motif finders are tools that identify such motifs in a given sequence; they can be very useful in finding about the function of a given sequence. MOTIF Search is one such tool which analyses the sequences that should be submitted in FASTA format, compares it with databases such as PROSITE and identifies the widely conserved motifs. Motif finders are used nowadays in a variety of studies such as genome wide identification of binding sites in various sequences (Kang *et al*., 2009).

## 2.4 Homology Modelling

Homology modelling is the process of modelling a structure for protein sequence, which doesn't have a recognized structure by comparing the sequence to other homologous proteins having known structures (Ginalski, 2006). It's a comparative modelling technique, which normally involves three steps, one, finding homologues in the Protein Data Bank (PDB), two, aligning the primary sequence with the retrieved protein sequences using multiple or single sequence alignments and three, calculating the structure and refining it (Zhang and Skolnick, 2005). Homology modelling techniques are very useful in finding conserved regions between sequences, in drug design studies and protein interaction studies. Homology modelling is based on the idea that structures of proteins that are slightly related have more observable conserved regions than the respective protein sequences (Renom et al., 2000). Homology modelling can be performed using various computational tools such as MODELLER, GeneSilico, and HHPred etc. Swiss-Model is a very useful reliable web server that can be used to create basic homology models. Homology modelling is widely used in bacterial protease studies, for example in a extensive study on the Pla protease of Yersinia pestis Swiss Model server has been used to create models (Suomalainen *et al*., 2007).

Of the three omptins under study, only OmpT of *E. coli* has an accepted structure at the PDB database (accession no. 1I78), PgtE of *Salmonella* and Pla of *Y. pestis* don't have any recognized protein structure annotated in the PDB repository. To study the conserved regions between PgtE and Pla of *Y. pestis* and OmpT of *E. coli*, the homology modelling technique was utilized to model tertiary protein structures for both PgtE and Pla using the OmpT structure as a template. The Swiss Model web server (http://swissmodel.expasy.org/), was used to model the crystal structures for the protein sequences of PgtE and Pla. The sequences in the FASTA format were submitted in the query space and the template structure required was selected from the PDB repository using the accession number mentioned for the crystal structure of OmpT. Structure validation and evaluation for the modelled structures were carried out using different techniques such as Ramachandran Plot and tools like ProCheck, Verify3D. A hypothesis on the possible functions of the proteinase PgtE that can be accessed from its structure was done using the ProFunc tool available in the EBI server (http://www.ebi.ac.uk/thornton-srv/databases/profunc/) and a complete set of PDBsum analysis on the modelled structure was also done using the same server (http://www.ebi.ac.uk/pdbsum/).

# 3. RESULTS AND DISCUSSION

## 3.1 PgtE

### 3.1.1 Sequence Alignment

The proteins sequence of PgtE obtained from the GenBank repository was run in an alignment search using BLAST. The sequences similar to that of PgtE and of those that were considered to be important to the study, because of the functional similarity they shared with PgtE, like ECM components cleavage, are listed in Table: 2. The list of proteases that contained sequence similarity to PgtE obtained through BLASTp has been listed in Appendix 1; the four proteases from Table 2 were selected for this study because they had been identified to have a certain level of functional similarity to PgtE through the research on published literature. In the four we know that the Pla of Yersinia pestis and OmpT of E. coli were from the same family of outer membrane proteases, the omptins. The other two proteases were further analysed among published literature to find correlation in their to that of PgtE. Through which we could determine a common structural motif for PgtE action on its substrates.

TABLE: 2

| Accession Number | Name of The Protein | Source Organism | Sequence Similarity Score |
|---|---|---|---|
| NP_995574.1 | Outer membrane protease (Pla) | Yersinia pestis biovar Microtus strain. 91001 | 484 |
| YP_001176628.1 | Outer membrane protease (OMP1) | Enterobactor clocae subsp. clocae ATCCC 13047 | 490 |
| NP_287408.1 | Outer membrane protease (OmpT) | Escherichia coli | 299 |
| NP_857613.1 | Outer membrane protease (OMP2) | Erwinia sp. Ejp 556 | 461 |

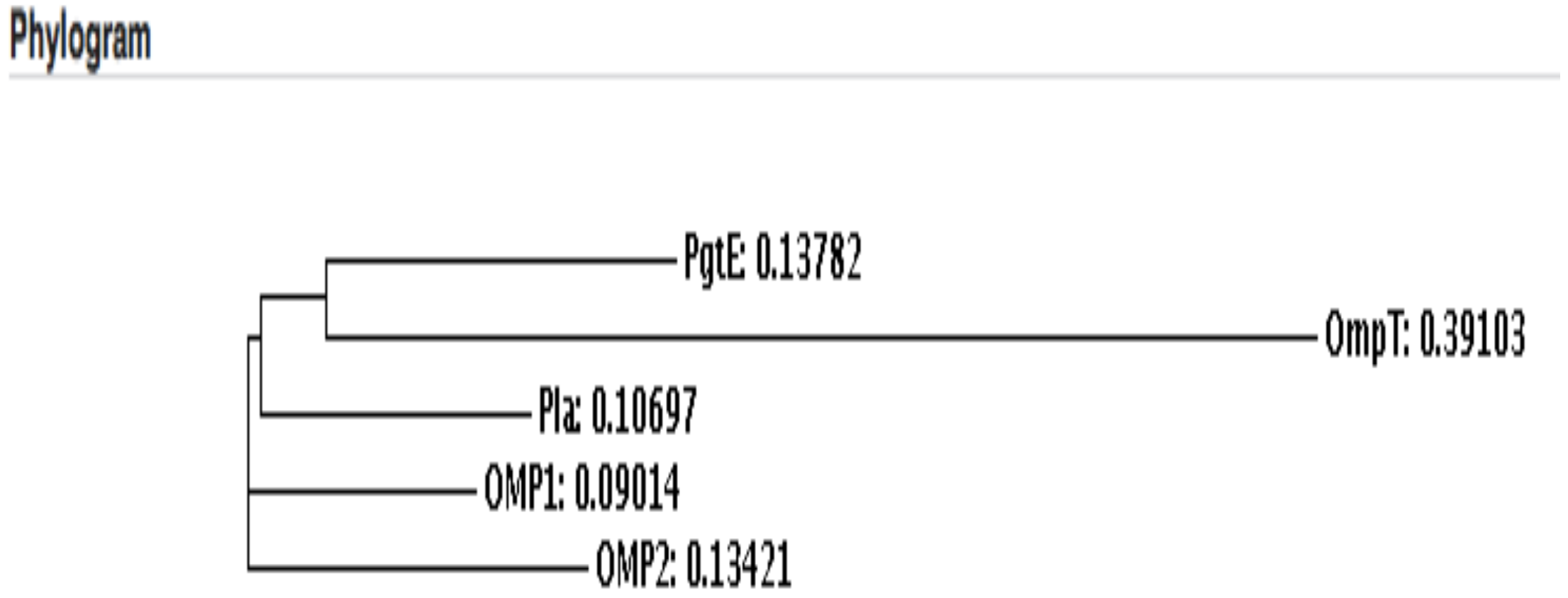

Fig 1: Shows the phylogenetic tree drawn for the four proteases obtained from the BLAST results along with PgtE. The evolutionary distance between them is also shown.

From the phylogenetic tree that was generated through the ClustalW analysis, the evolutionary divergence and distance between the different proteases used in that experiment could be observed. This phylogram shows that, PgtE, Pla and OmpT have a common ancestor, which is understandable because they belong to the same category of proteases called the omptins. It can be also seen from the evolutionary distance calculated that PgtE and Pla had diverged recently though OmpT is considerably a far distance off. The other proteases OMP1 and OMP2, though they have sequence similarity to PgtE, they are far distanced in evolutionary relation. It's understandable because they are both from different protease families entirely. The significance of this result is that it has been known that while evolving PgtE and the omptins had developed different modifications to their surface loops, which differentiates their functions. The omptin family presents a highly interesting group of evolutionarily related proteases that have evolved to fit into the life style of their host bacteria and prove how the β-barrel through the evolutionary process can be modified functionally by small changes at critical exposed sites on the surface (Kukkonen and Korhonen, 2004).

ClustalW Results

The next step on the analysis of PgtE involved the multiple sequence alignment of PgtE with all the proteases from Table 2 using ClustalW.

```
OMP1            MTKNIVAGMMMAAFSGAVYAGSAQAIPDFSPESLSVSASAGMLSGKSHEMVYDEVTG-RK 59
OMP2            MKKSTVLVTMMSAFFGTDYVAAAQISPDFSADSMSVATSVGILGGKSKELVYDASNA-RK 59
Pla             MKKSSIVATIITILSGSANAASSQLIPNISPDSFTVAASTGMLSGKSHEMLYDAETG-RK 59
PgtE            MKKHAIAVMMIAVFSESVYAESALFIPDVSPDSVTTSLSVGVLNGKSRELVYDTDTG-RK 59
OmpT            MRAKLLGIVLTTPIAISSFASTETLS--FTPDNINADISLGTLSGKTKERVYLAEEGGRK 58
                *    :   : : :  :  . :      .:.:....  * * *.**::* :*     . **

OMP1            ISQLDWKIKNVAILKGDISWDAYSFLTLNARGWTSLASGSGHMDDYDWM-NAKQSSWTDH 118
OMP2            ISQLDWKIKNVAIIKADFSWDAYSFLTLNARGWTSLASGSGHMDDYDWQ-NSNQSDWTDH 118
Pla             ISQLDWKIKNVAILKGDISWDPYSFLTLNARGWTSLASGSGNMDDYDWM-NENQSEWTDH 118
PgtE            LSQLDWKIKNVATLQGDLSWEPYSFMTLDARGWTSLASGSGHMVDHDWM-SSEQPGWTDR 118
OmpT            VSQLDWKFNNAAIIKGAINWDLMPQISIGAAGWTTLGSRGGNMVDQDWMDSSNPGTWTDE 118
                :******::*.* ::. :.*:  . :::.* ***:*.* .*:* * **  . :   ***.

OMP1            SSHPATNVNYANEYDLNVKGWIFQGDNYKAGVTAGYQETRFSWTATGGSYNYDNGT---- 174
OMP2            SSHPGTDVNYANEYDLNLKGWFLQGNDYKVGAVAGYQETRFSWTATGGSYSYNNGA---- 174
Pla             SSHPATNVNHANEYDLNVKGWLLQDENYKAGITAGYQETRFSWTATGGSYSYNNGA---- 174
PgtE            SIHPDTSVNYANEYDLNVKGWLLQGDNYKAGVTAGYQETRFSWTARGGSYIYDNGR---- 174
OmpT            SRHPDTQLNYANEFDLNIKGWLLNEPNYRLGLMAGYQESRYSFTARGGSYIYSSEEGFRD 178
                * ** *.:*:***:***:***:::  :*: *  *****:*:*:** **** *..

OMP1            NTGNFPAGERGIGYSQRFSMPYIGLAGQYRFNDFEFNALFKFSDWVRAHDNDEHY--MRD 232
OMP2            SVGNFPNQRPGIGYSQRFSMPYIGLVGQYRINDFEFNALFKFSDWVRAHDNDEHY--MRS 232
Pla             YTGNFPKGVRVIGYNQRFSMPYIGLAGQYRINDFELNALFKFSDWVRAHDNDEHY--MRD 232
PgtE            YIGNFPHGVRGIGYSQRFEMPYIGLAGDYRINDFECNVLFKYSDWVNAHDNDEHY--MRK 232
OmpT            DIGSFPNGERAIGYKQRFKMPYIGLTGSYRYEDFELGGTFKYSGWVEASDNDEHYDPGKR 238
                  *.**     ***.***.******.*.** :***  .  **:*.**.* ******   :

OMP1            LTFREKTSDSRYYGASVDAGYYVTPNAKVFAEFTYSSYEEGKGGTQIIDTNTGESGSIGG 292
OMP2            LTFREKTSDSRYYGASVDAGYYVTRNAKVFAEFSYSKYEEGKGGTQIIDTISGDSASLDG 292
Pla             LTFREKTSGSRYYGTVINAGYYVTPNAKVFAEFTYSKYDEGKGGTQIIDKNSGDSVSIGG 292
PgtE            LTFREKTENSRYYGASIDAGYYITSNAKIFAEFAYSKYEEGKGGTQIIDKTSGDTAYFGG 292
OmpT            ITYRSKVKDQNYYSVSVNAGYYVTPNAKVYVEGTWNRVTNKKGNTSLYDHN-DNTSDYSK 297
                :*:*.*.....**.. ::****:* ***::.* ::.   : **.*.: *   .::   .

OMP1            DAAGISNRNYTITAGLQYRF 312
OMP2            DAAGISNKNYTATVGLQYRF 312
Pla             DAAGISNKNYTVTAGLQYRF 312
PgtE            DAAGIANNNYTVTAGLQYRF 312
OmpT            NGAGIENYNFITTAGLKYTF 317
                :.*** * *:  *.**:* *
```

Fig 2: Shows the ClustalW alignment of five proteases- PgtE of Salmonella, Pla of Y. pestis, OmpT of E.coli and the outer membrane proteases of both E. clocae and Erwinia sp. '*' represent the conserved regions among the sequences.

From analysing the above alignment and the alignment scores obtained through the ClustalW tool, it is clear that all these sequences possess considerable sequence similarity as well as the presence of conserved regions could be clearly identified, especially at the C- terminal and throughout the sequences except after the start from the N-terminal. There are a few gaps, which maybe the results of mutations during evolution.

### 3.1.2 Motif Identification

MOTIF Search tool was used to find the already recognised motifs among the PgtE protein sequence by comparing with the PROSITE database. Three known motifs were found in PgtE, two Aspartyl protease omptin family signatures and an EF- hand Calcium binding domain. They are illustrated in Table: 3 below.

| FOUND MOTIF | POSITION | PROSITE ID | DESCRIPTION |
|---|---|---|---|
| WTDRSIHPDT | 115-124 | OMPTIN_1 (PSOO834) | Aspartyl protease, omptin family signature 1. |
| AGYQETRFSWTARGGSY | 152-168 | OMPTIN_2 (PS00835) | Aspartyl protease, omptin family signature 2. |
| DTDTGRKLSQLDW | 53-65 | EF_HAND_1 (PS00018) | EF-hand calcium binding domain. |

### 3.1.3 Homology Modelling

To study the conserved domains using the structural data of the protein sequences, the crystal structure of PgtE was modelled using Swiss-Model homology modelling tool, using OmpT of E.coli as a valid template, since they share more that 48% sequence similarity. The structure of Pla of Y. pestis was also modelled using the same process. Both these modelled structures were super imposed on each other and were allowed to undergo structure alignment to study the differences in their loops, which corresponds to their specific activities.

The modelled structures of PgtE and Pla shows a high range of similarity, they both possess the characteristic cylindrical barrel shape of the omptins, much like OmpT of E. coli. In both the structures there are 5 (L1 - L5) exposed loops on the surface and ten transmembrane beta strands, this can be seen normally within all of the omptins as well (Rutten et al., 2001;Kukkonen et al., 2001). The transmembrane beta strands can undergo large substitutions in their exposed areas of the surface, since they are very stable (Wimley, 2003).

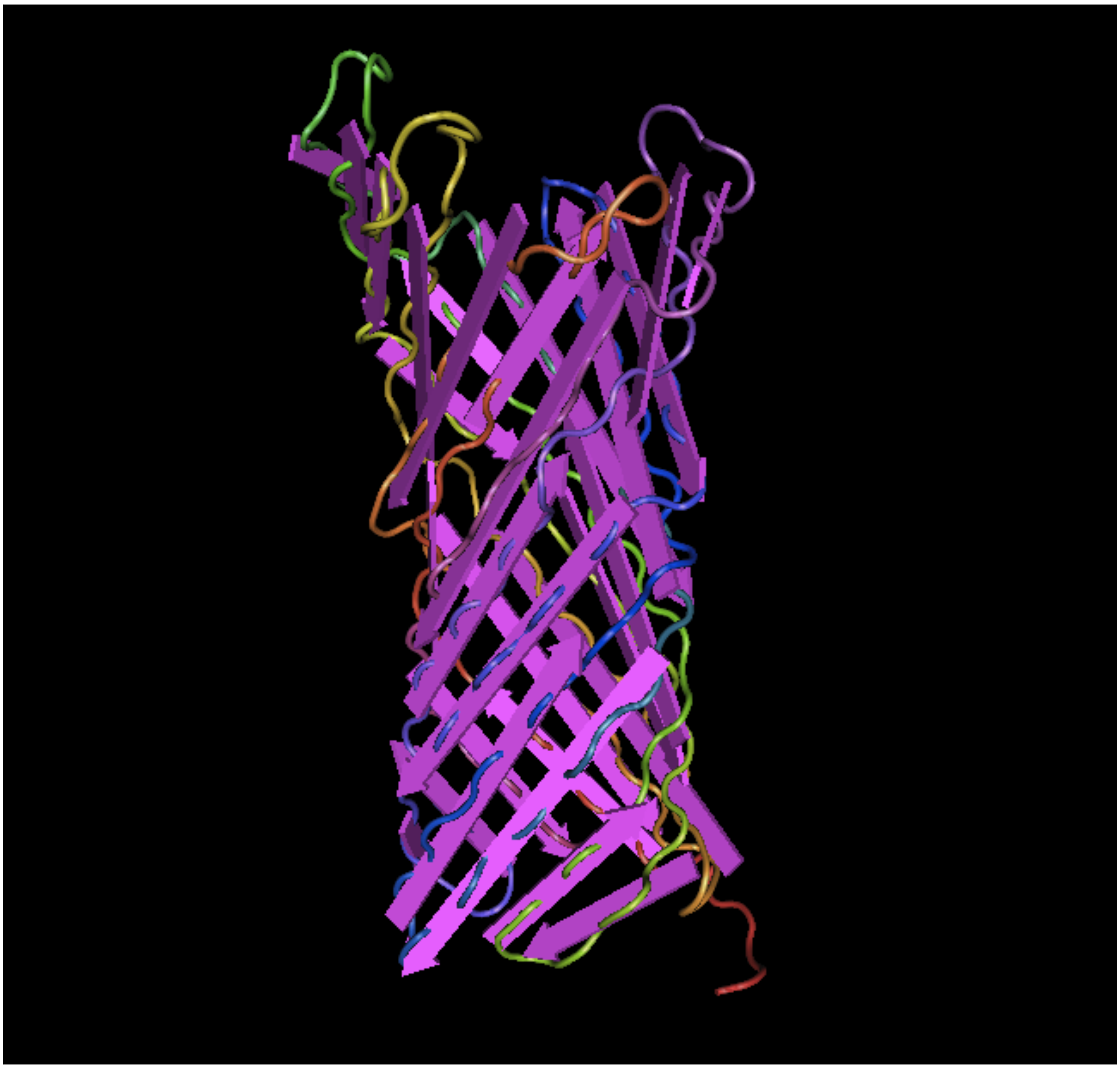

Fig 3: Jmol image showing the cylindrical structure of the modelled PgtE of Salmonella enterica serovar Typhimurium.

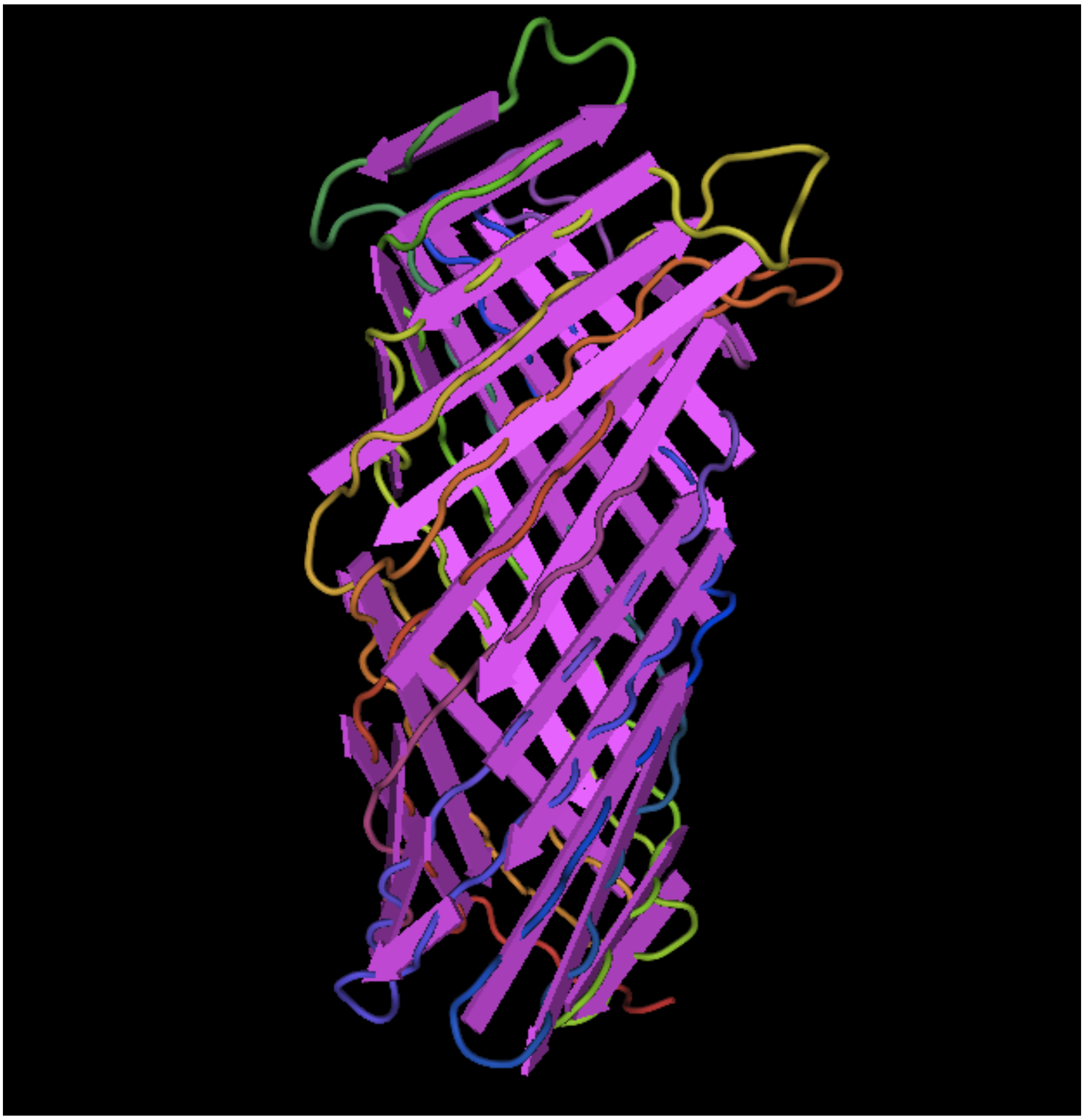

Fig 4: Jmol image showing the cylindrical structure of the modelled Pla of Y. pestis.

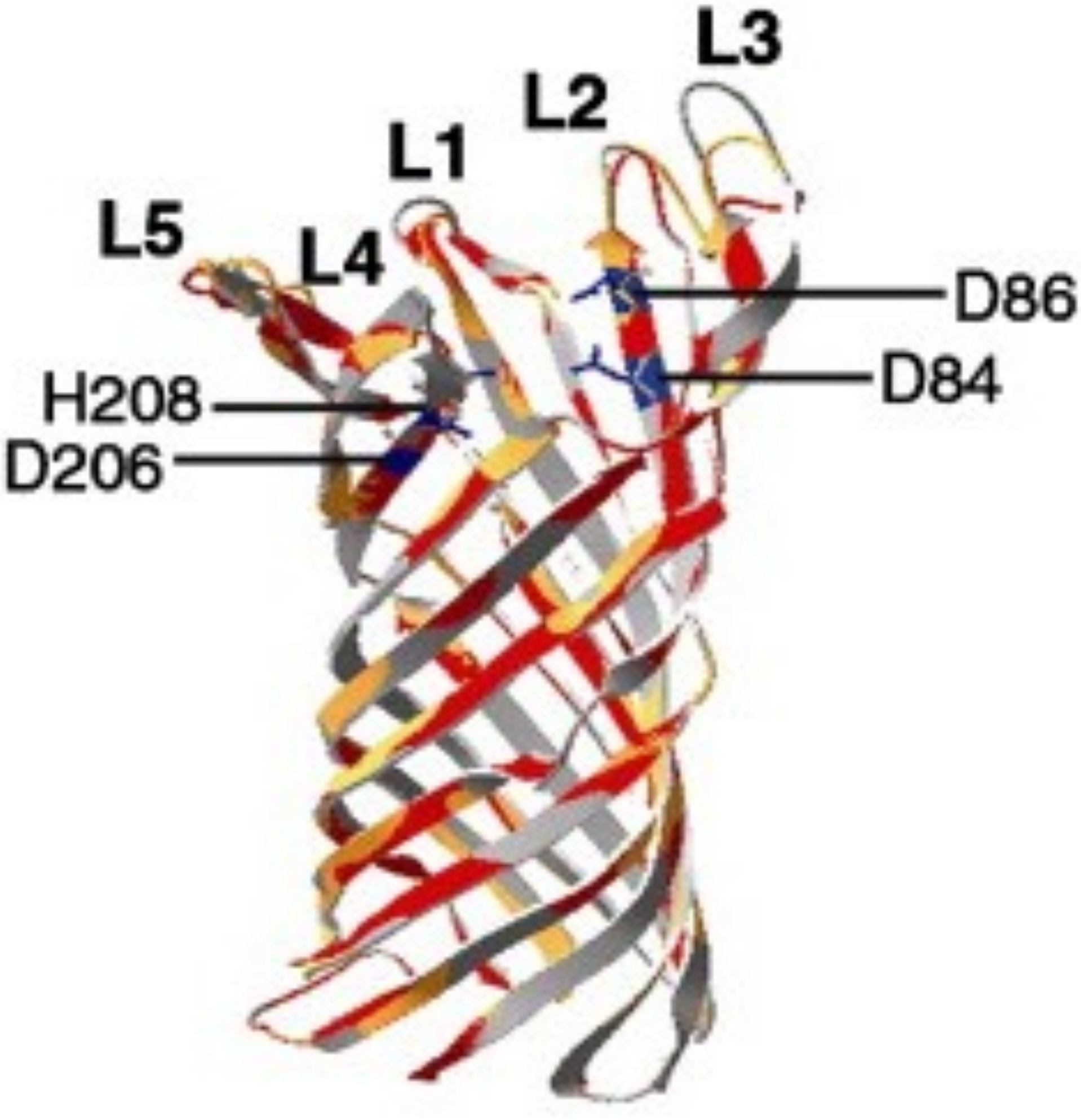

Fig 5: A model showing the different loops and important surface residues in superimposed structure of PgtE and Pla. (Adapted from Ramu *et al*., 2007)

The modelled structures of PgtE and Pla shows a high range of similarity, they both possess the characteristic cylindrical barrel shape of the omptins, much like OmpT of E. coli.  In both the structures there are 5 (L1 - L5) exposed loops on the surface and ten transmembrane beta strands, this can be seen normally within all of the omptins as well (Rutten et al., 2001;Kukkonen et al., 2001). The transmembrane beta strands can undergo large substitutions in their exposed areas of the surface, since they are very stable (Wimley, 2003). These substitutions could be used as a way of converting or adding properties to the omptins, which they didn’t have in the wild condition. For example, studies have shown that experiments have successfully converted Pla into a PgtE like sequence, by performing simple substitutions, which further allowed Pla to have PgtE like functions, such as activating proMMP-9 (Ramu *et al*., 2007). These substitution regions and loops are shown clearly in Figure 5.

Further the binding sites of PgtE were analysed through PDBsum in the EBI web server, the position of the binding sites and the residues in it were identified.

Residue type colouring scheme is given below,

Red – Negative Residues (D, E).

Blue – Positive Residues (H, K, R).

Purple – Aromatic Residues (F, Y, W).

Yellow – Cysteine (C).

These are the binding sites that are normally involved in the catalytic activity of PgtE.

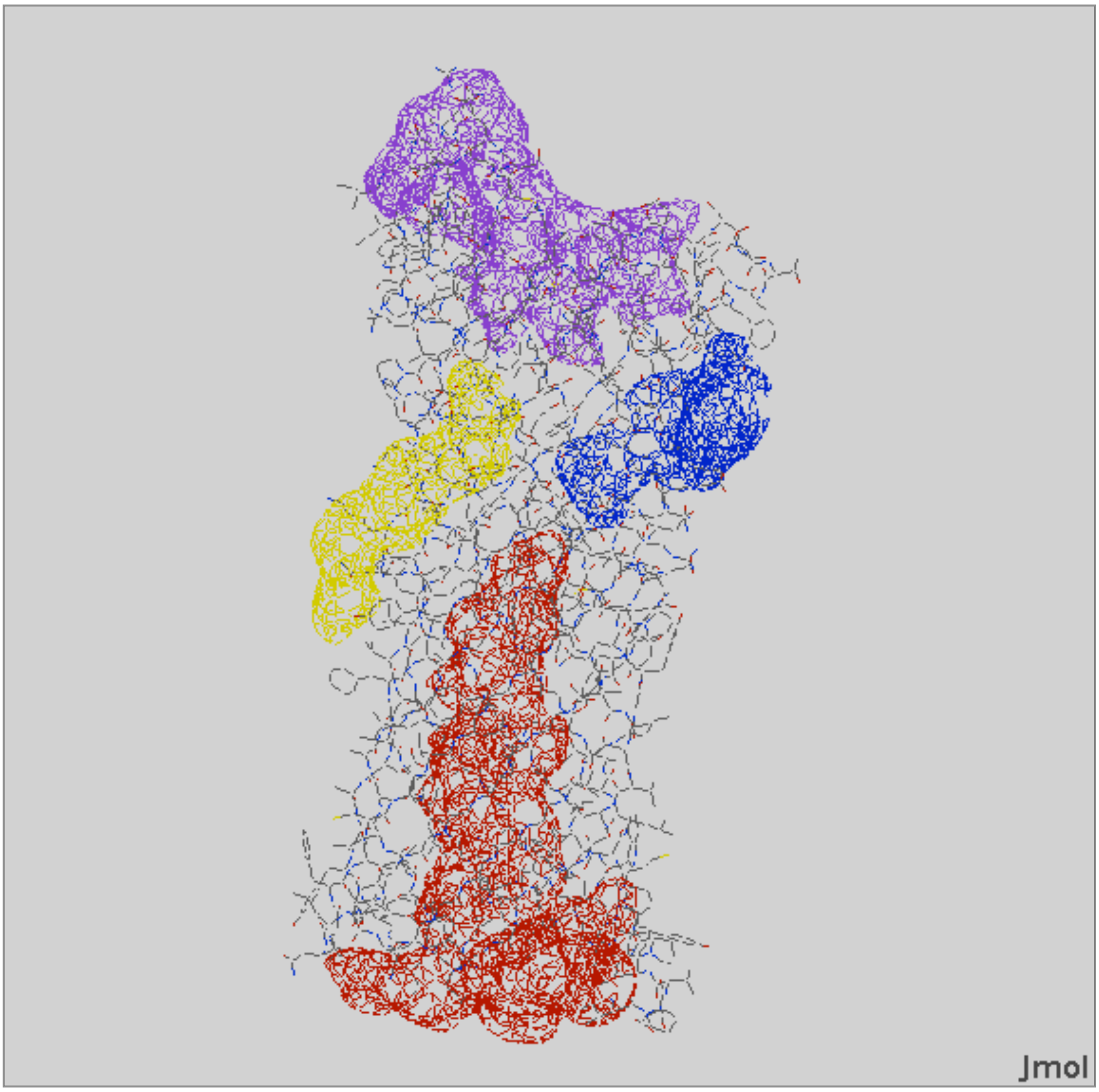

Fig 6: Shows the different general binding sites observer in PgtE in colour generated by PDBsum, each colour corresponds to the different residues involved.

3.2 PrtA

The protein sequences of the substrates in which PrtA was known to act was analysed through a refined study of published literature. It was observed to be active against 16 proteins in *Manduca sexta* of which 6 where known to have some kind of immunological activity. The other 10 were unknown proteins with no clarity on their function. These proteins also known as PAT proteins are listed in Table 4 below.

| PAT PROTEIN | NAME & ACCESSION NO. | HYPOTHESIZED FUNCTION |
|---|---|---|
| PAT - 35a | Scolexin A (AAD14591.1) | Immunological |
| PAT - 35b | Scolexin B (AAD14592.1) | Immunological |
| PAT - 41a | SPH-3 (P13082.1) | Immunological |
| PAT - 52 | Serpin-1 (AAC47343.1) | Immunological |
| PAT - 54a | HAIP (ACW82749.1) | Immunological |
| PAT - 63 | β1,3-GRP-2 (AAN10151.1) | Immunological |

All the six PAT proteins were analysed using multiple sequence alignment to identify conserved regions that might indicate some kind of substrate specificity by PrtA, ClustalW from the EBI server was used for this process.

ClustalW Result

The multiple sequence alignment of the PAT proteins showed that they don't share any considerable sequence similarity, though all of them shared some immune related function. A phylogenetic tree was built was also built which showed the considerable evolutionary distance between the different PAT proteins and how they have diverged.

```
ScolexinA       ------RQSVVLAVAAVLFGCAC------------------------AAPNPGANDIQLN 30
ScolexinB       -----SKQSVVLAVAAALVACAC------------------------AAPDPGANDIQLN 31
Î²1_3-GRP-2     ----MLKSVFVLFLVNYLNSVRCLEVPDAK------LEAIYPKGLRVSIPDDGYTLFAFH 50
SPH-3           -----MERWRLLRDGELLTTHSS---------------------WILPVRQGDMPAMLKV 34
HAIP            MKILIALAGVLAVAVATPASVPRKVLCYYD------SKSYVRESQARMLPMDLDPALSFC 54
Serpin-1        MKIIMCIFGLAALAMAGETDLQKILRESNDQFTAQMFSEVVKANPGQNVVLSAFSVLPPL 60
                                                                        :

ScolexinA       QKLSIEAKGAKQPIDTR------------------AVKERYPYAVRSFGGFCGG-----T 67
ScolexinB       QKLSVDAKGAKQPIDTR------------------AVNERYPHAVL-FGGTCGG-----T 67
Î²1_3-GRP-2     GKLNEEMEGLEAGHWSRDIT--------------KAKNGRWIFRDRNAKLKIGD-----K 91
SPH-3           ARIPDEEAGYRLLTWWD------------------------GQGAARVFASAAG-----A 65
HAIP            THLLYGYAGIQPDTYKMVPLNENLDVDRAHANYRAITNFKTKYPGLKVLLSVGGDADTEE 114
Serpin-1        GQLALASVGESHDELLR------------------ALALPNDNVTKDVFADLNRGVRAVK 102
                 ::     *

ScolexinA       IISPTWILTAGHCSILYAGSGLPAGTNITE-----VSSLYR-FPKRLVIHPLFSIG---- 117
ScolexinB       IISPTWILTAGHCTLFNDGRGVLAGTNNSD-----VSGVYR-FTKRLIIHPLFSVG---- 117
Î²1_3-GRP-2     IYFWTYILKDGLGYRQDNGEWTVTGYVNEDGEPLDANFEPR-STASTAAPPQAGAGQAPG 150
SPH-3           LLMERASGAGDLAQIAWSGQDDEACRILCDT--AARLHAPR-SGPPPDLHPLQEWFQP-- 120
HAIP            AQKYNLLLESPQARTAFVNSGVLLAEQHGFDGIDLAWQFPR-IKPKKVRSTWGSIWHG-- 171
Serpin-1        GVDLKMASKIYVAKGLELNDDFAAVSRDVFGSEVQNVDFVKSVEAAGAINKWVEDQTNNR 162
                                  .                     :

ScolexinA       ---PVWLNATEFNLKQAAARWDFLLIELEEPLPLDGKILAAAKLD-DQPDLPA------- 166
ScolexinB       ---PYWLNAEEFNLKQVAARWDFLLAELEEPLPLDGKIMAAAKLD-DQPDLPA------- 166
Î²1_3-GRP-2     PSYPCELSVSEVSVPGFVCKGQMLFEDNFNKPLADGRIWTPEIMFPGEPDYPFNVYMKET 210
SPH-3           ---LFRLAAEHAALAPAASVARQLLAAPREVCPLHGDLHHENVLDFGDRGWLAIDPHG-- 175
HAIP            IKKTFGTTPVDDKEAEHREGFTALVRELKQALNVKPNMQLAVTVLPNVNASIYYDVPAII 231
Serpin-1        IKNLVDPDALDETTRSVLVNAIYFKGSWKDKFVKERTMDRDFHVSKDKTIKVPTMIGKKD 222
                          .            :     :    .  :     :  .

ScolexinA       -GLDVG-----------------------------------YPSYSTDTYEAKIQSEMHG 190
ScolexinB       -GLDVG-----------------------------------YAGYGTDHHGGTMRSEMHA 190
Î²1_3-GRP-2     DNLHVGNGNLVIKPMPLVTAFGEDAIWKTLDLSDRCTGLLGTAQCKRDPSDAIIVPPIVT 270
SPH-3           ----------------------------------------LLGERTFDYANIFTNPDLSD 195
HAIP            NLVDIVN--------------------IEAYDYFTPERNPKEADYVSPIYTPQNRNPLQN 271
Serpin-1        VRYADVP--------------------------------ELDAKMIEMSYEGDQASMIII 250
                                                                         :
```

```
ScolexinA        KKLSVQS----------------------------------------------------- 197
ScolexinB        MELSVQS----------------------------------------------------- 197
Î²1_3-GRP-2      AKINTKKTFAFKYGRVEISAKMPRGDWLVPLIQLEPVNKNYGIRNYVSGLLRVACVKGNT 330
SPH-3            PGRPLA------------------------------------------------------ 201
HAIP             VDAAINYWLQSN------------------------------------------------ 283
Serpin-1         LPNQVDG----------------------------------------------------- 257


ScolexinA        --------NEVCSKLEQFKAEDMLCAKGRPPR--------------YDFVCFSDSGSGLV 235
ScolexinB        --------NEVCSKLEQFEAKDMLCAKGRPPR--------------YDSACNGDSGSGLV 235
Î²1_3-GRP-2      EYIKTLVGGPIMSEAEPYRTANLKEFISNEPW--TNE------FHNYTLEWSPDAITMAV 382
SPH-3            ----------ILPGRLEARLSIVVATTGFEPERL----------LRWIIAWTGLSA---- 237
HAIP             -APSNKLVLGIASYSRTWKLDSESEISGVPPLHTDGAGEAGPYTKIEGLLSYPEVCAKLI 342
Serpin-1         ---ITALEQKLKDPKALSRAEERLYNTEVEIYLP-----------KFKIETTTDLKEVLS 303
                           :       .

ScolexinA        D------------------NNGRLVGVVSWAENN-----------AFECRNGNLAVFSRV 266
ScolexinB        D------------------NNGRLVGVASWVEND-----------AFECRNGNLVVFSRV 266
Î²1_3-GRP-2      DGIVYGRVTAPAGGFYKEANEQNVEAAARWIQGSNIAPFDDMFYISLGMDVGGVHEFPDE 442
SPH-3            ----------------------------AWFIGD-----------GDGEGEGAAIDLAVN 258
HAIP             N---------------PNHQKGMRPHLRKVTDPSKRFGTYAFRLPDDSGEGGMWVSYEDP 387
Serpin-1         N---------------MNIKKLFTPGAARLENLLKTKES----LYVDAAIQKAFIEVNEE 344


ScolexinA        SSVREWIRQVTNI----------------------------------- 279
ScolexinB        SSVREWIRQVTNI----------------------------------- 279
Î²1_3-GRP-2      AINKPWKNTATKAMVNFWNARSQWNP-TWLESEKALLVDYVRVYAL-- 487
SPH-3            AMARRLLD---------------------------------------- 266
HAIP             DTAGQKASYVTSKNLGGISINDLSMD-DFRGLCTGDKYPILRAAKYRL 434
Serpin-1         GAEAAAANAFGIVPASLILYPEVHIDRPFYFELKIDGIPMFNGKVIEP 392
```

Fig 7: Shows the ClustalW alignment between the six PAT proteins on which PrtA acts.'*' Represents the conserved regions among the sequences.

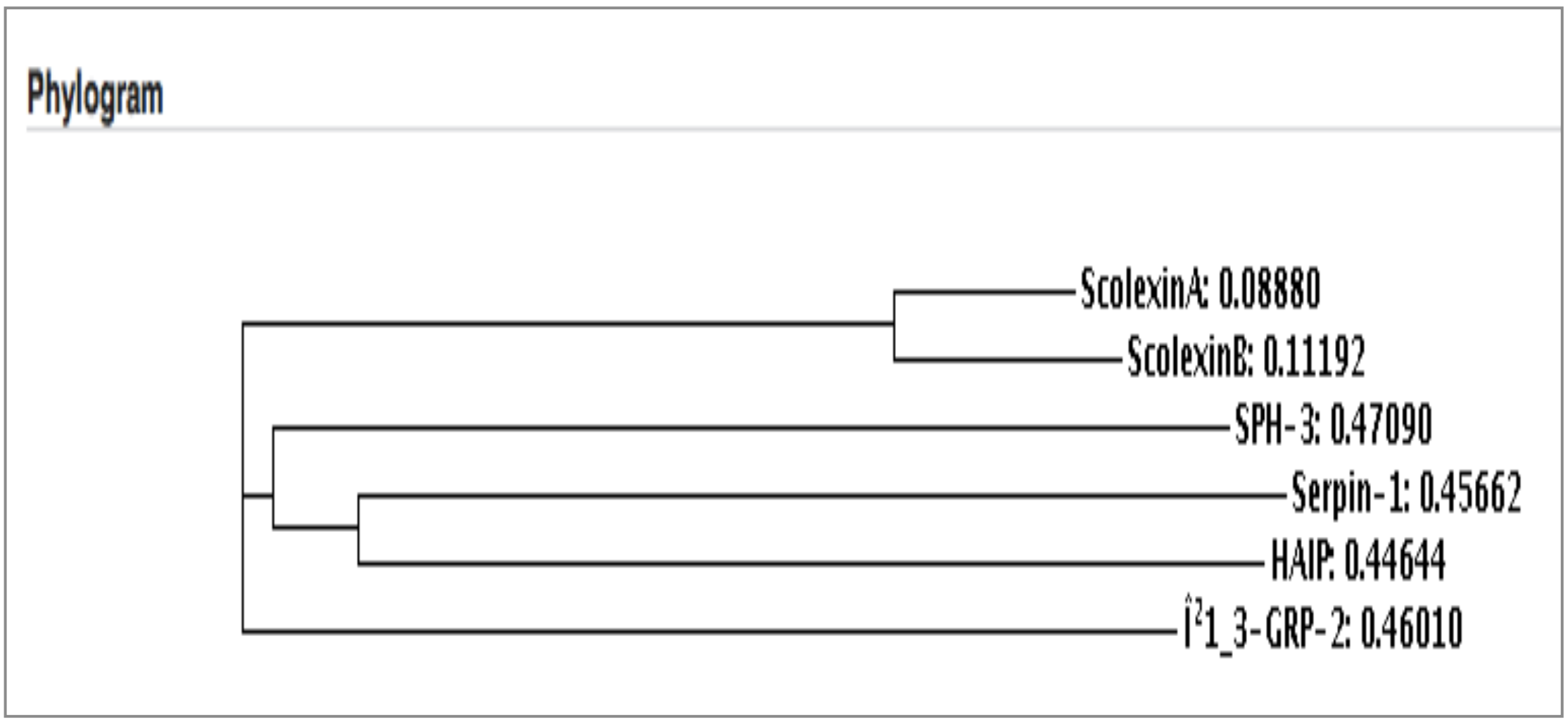

Fig 8: Phylogenetic Tree showing the evolutionary divergence and distance between the six PAT proteins on which PrtA acts.

From the phylogram it can be observed that Scolexin A and B have a common ancestor which is irrefutable because they are just variants of the same protease. Also an interesting thing was that HAIP and Serpin- 1 share common ancestry with recent divergence, though there is not much sequence similarity. SPH-3 also shares a common ancestor with them, though it has diverged earlier.

3.3 GINGIPAINS

The protein sequences of the gingipains RgpA and Kgp of P. gingivalis was obtained from the Genbank repository through the NCBI web server. They were subjected to multiple sequence alignment using ClustaW through the EBI web server.

```
Rgp                 MNKFVSIALCSSLLGGMAFAQQTELGRNPNVRLLESTQQSVT--KVQFRMDNLKFTEVQT 58
Kgp                 MRKLL-LLIAASLLGVGLYAQSAKIKLDAPTTRTTCTNNSFKQFDASFSFNEVELTKVET 59
                    *.*:: : :.:****   :**.:::  :. .    .*::*..  ...* ::::::*:*:*

Rgp                 PKGMAQVPTYTEGVNLSEKGMPTLPILSRSLAVSD--TREMKVEVVSSKFIEK---KNVL 113
Kgp                 KGGTFASVSIPGAFPTGEVGSPEVPAVRKLIAVPVGATPVVRVKSFTEQVYSLNQYGSEK 119
                      *     : . ..  .* * * :* : : :**.   *  ::*: .:.:. .     .

Rgp                 IAPSKGMIMRNEDPKKIPYVY-GKSYSQNKFFPGEIATLDDPFILRDVRGQVVNFAPLQY 172
Kgp                 LMPHQPSMSKSDDPEKVPFVYNAAAYARKGFVGQELTQVEMLGTMRGVRIAALTINPVQY 179
                    : * :  : :.:**:*:*:** . :*::: *.  *:: ::    :*.**  .:.: *:**

Rgp                 NPVTKTLRIYT--EITVAVSETSEQGKNILNKKGTFAGFEDTYKRMFMNYEPGRYTPVEE 230
Kgp                 DVVANQLKVRNNIEIEVSFQGADEVATQRLYDASFSPYFETAYKQLFN---RDVYTDHGD 236
                    : *:: *:: .  ** *:.. :.* ..: * . .  . ** :**::*     . **   :

Rgp                 KQNG--RMIVIVAKKYEGDIKDFVDWKNQRGLRTEVKVAEDIASPVTANAIQQFVKQEYE 288
Kgp                 LYNTPVRMLVVAGAKFKEALKPWLTWKAQKGFYLDVHYTDEAEVGTTNASIKAFIHKKYN 296
                      *   **:*:.. *::  :* :: ** *:*:  :*: :::    .*  :*: *::::*:

Rgp                 K---EGNDLTYVLLVGDHKDIP-AKITPGIKSDQVYGQIVGNDHYNEVFIGRFSCESKED 344
Kgp                 DGLAASAAPVFLALVGDTDVISGEKGKKTKKVTDLYYSAVDGDYFPEMYTFRMSASSPEE 356
                    .    .   .:: **** . *.  * .   *  ::* . *..*:: *::  *:*..* *:

Rgp                 LKTQIDRTIHYERNITTEDKWLGQALCIASAEGGPSADNGESDIQHENVIANLLTQYGYT 404
Kgp                 LTNIIDKVLMYEKATMPDKSYLEKALLIAGADSYWNPKIGQQTIKYA-VQYYYNQDHGYT 415
                    *.. **:.: **:   .:..:* :** **.*:.  ... *:. *::  *      ::***

Rgp                 KIIKCYDPGVTPKNIIDAFNGGISLVNYTGHGSETAWGTSHFGTTHVKQLTNSNQLPFIF 464
Kgp                 DVYSY--PKAPYTGCYSHLNTGVGFANYTAHGSETSWADPSLTATQVKALTNKDKYFLAI 473
                    .: .   * .. ..  . :* *:.:.***.*****:*. . : :*:** ***.::  : :

Rgp                 DVACVNGDFLFSMPCFAEALMRAQKDGKPTGTVAIIASTINQSWASPMRGQDEMNEILCE 524
Kgp                 GNCCVTAQFDYPQPCFGEVMTRVKE----KGAYAYIGSSPNSYWGEDYYWSVGANAVFGV 529
                    . .**..:* :. ***.*.: *.::    .*: * *.*: *. *..    .   * ::

Rgp                 KHPNN--IKRTFGGVTMNGMFAMVEKYKKDG-----------------EKMLDTWTVFG 564
Kgp                 QPTFEGTSMGSYDATFLEDSYNTVNSIMWAGNLAATHAGNIGNITHIGAHYYWEAYHVLG 589
                    : . :     ::... ::. :  *:.    *                 .   ::: *:*
```

```
Rgp             DPSLLVRTLVPTKMQVTAPAQINLTDASVNVSCDYNGAIATISANGKMFGSAVVEN-GTA 623
Kgp             DGSVMPYRAMPKTNTYTLPASLPQNQASYSIQAS-AGSYVAISKDGVLYGTGVANASGVA 648
                * *::    :*..   * **.:  .:** .:...  *: .:** :* ::*:.*.:  *.*

Rgp             TINLTGLTNES-TLTLTVVGYNKETVIKTINTNGEPNPYQPVSNLTATTQGQKVTLKWDA 682
Kgp             TVNMTKQITENGNYDVVITRSNYLPVIKQIQA-GEPSPYQPVSNLTATTQGQKVTLKWDA 707
                *:*:*   .*. .  :.:.  *  .*** *:: ***.***********************

Rgp             PSTKTNATTNTARSVDGIRELVLLSVSDAPELLRSGQAEIVLEAHDVWNDGSGYQILLDA 742
Kgp             PSAKKAEAS---REVKRIGDGLFVTIEPANDVR-ANEAKVVLAADNVWGDNTGYQFLLDA 763
                **:*.  ::   *.*. * : :::::. * ::  :.:*::** *.:**.*.:***:****

Rgp             DHDQYGQVIPSDTHTLWPNCSVPANLFAPFEYTVPENADPSCSPTNMIMDGTASVNIPAG 802
Kgp             DHNTFGSVIP-ATGPLFTGTASSNLYSANFEYLIPANADPVVTTQNIIVTGQGEVVIPGG 822
                **: :*.***  * .*:.. : .    * *** :* ****  :. *:*: * ..* **.*

Rgp             TYDFAIAAP-QANAKIWIAGQG---PTKEDDYVFEAGKKYHFLMKKMGSGDGTELTISEG 858
Kgp             VYDYCITNPEPASGKMWIAGDGGNQPARYDDFTFEAGKKYTFTMRRAGMGDGTDMEVEDD 882
                .**:.*: *  *..*:****:*   *:: **:.******* * *:: * ****:: :.:.

Rgp             GGSDYTYTVYRDGTKIKEGLTATTFEEDGVATGNHEYCVEVKYTAGVSPKVCKDVTVEGS 918
Kgp             SPASYTYTVYRDGTKIQEGLTATTFEEDGVAAGNHEYCVEVKYTAGVSPKVCKDVTVEGS 942
                . :.************:**************:****************************

Rgp             NEFAPVQNLTGSAVGQKVTLKWDAPNGTPNPNPNPNPNPNPGTTTLSESFENGIPASWKT 978
Kgp             NEFAPVQNLTGSAVGQKVTLKWDAPNGTPNPNPNP----NPGTTTLSESFENGIPASWKT 998
                ***********************************    *********************

Rgp             IDADGDGHGWKPGNAPGIAGYNSNGCVYSESFGLGGIGVLTPDNYLITPALDLPNGGKLT 1038
Kgp             IDADGDGHGWKPGNAPGIAGYNSNGCVYSESFGLGGIGVLTPDNYLITPALDLPNGGKLT 1058
                ************************************************************

Rgp             FWVCAQDANYASEHYAVYASSTGNDASNFTNALLEETITAKGVRSPEAIRGRIQGTWRQK 1098
Kgp             FWVCAQDANYASEHYAVYASSTGNDASNFTNALLEETITAKGVRSPEAIRGRIQGTWRQK 1118
                ************************************************************

Rgp             TVDLPAGTKYVAFRHFQSTDMFYIDLDEVEIKANGKRADFTETFESSTHGEAPAEWTTID 1158
Kgp             TVDLPAGTKYVAFRHFQSTDMFYIDLDEVEIKANGKRADFTETFESSTHGEAPAEWTTID 1178
                ************************************************************
```

```
Rgp     ADGDGQGWLCLSSGQLDWLTAHGGTNVVASFSWNGMALNPDNYLISKDVTGATKVKYYYA 1218
Kgp     ADGDGQDWLCLSSGQLDWLTAHGGTNVVASFSWNGMALNPDNYLISKDVTGATKVKYYYA 1238
        ******.*****************************************************

Rgp     VNDGFPGDHYAVMISKTGTNAGDFTVVFEETPNGINKGGARFGLSTEANGAKPQSVWIER 1278
Kgp     VNDGFPGDHYAVMISKTGTNAGDFTVVFEETPNGINKGGARFGLSTEANGAKPQSVWIER 1298
        ************************************************************

Rgp     TVDLPAGTKYVAFRHYNCSDLNYILLDDIQFTMGGSPTPTDYTYTVYRDGTKIKEGLTET 1338
Kgp     TVDLPAGTKYVAFRHYNCSDLNYILLDDIQFTMGGSPTPTDYTYTVYRDGTKIKEGLTET 1358
        ************************************************************

Rgp     TFEEDGVATGNHEYCVEVKYTAGVSPKECVNVTINPTQFNPVKNLKAQPDGGDVVLKWEA 1398
Kgp     TFEEDGVATGNHEYCVEVKYTAGVSPKVCVNVTINPTQFNPVKNLKAQPDGGDVVLKWEA 1418
        *************************** ********************************

Rgp     PSAKKTEGSREVKRIGDGLFVTIEPANDVRANEAKVVLAADNVWGDNTGYQFLLDADHNT 1458
Kgp     PSGKRGELLN--EDFEGDAIPTGWTALDADGDGNNWDITLNEFTRGERHVLSPLRASN-- 1474
        **.*: *  .  : : .. : *  .* *. .:  :  :: ::.  .:      * *.:

Rgp     FGSVIPATGPLFTGTASSNLYSANFEYLIPANADPVVTTQNIIVTGQGEVVIPGGVYDYC 1518
Kgp     ---VAISYSSLLQGQEYLPLTPNNFLITPKVEGAKKITYK---VGSPGLPQWSHDHYALC 1528
           *  : ..*: *     * . **     .:.   :* :   * . *    . . *  *

Rgp     ITNPEPASG-------KMWIAGDGGNQPARYDDF----TFEAGKKYTFTMRRAGMGDGTD 1567
Kgp     ISKSGTAAADFEVIFEETMTYTQGGANLTREKDLPAGTKYVAFRHYNCTDVLGIMIDDVV 1588
        *::. .*:.       :     :** : :* .*:    .: * ::*. *   . * *..

Rgp     MEVEDDSPASYTYTVYRDGTKIKEGLTETTYRDAGMSAQSHEYCVEVKYAAGVSPKVCVD 1627
Kgp     ITGEGEGPS-YTYTVYRDGTKIQEGLTETTYRDAGMSAQSHEYCVEVKYAAGVSPKVCVD 1647
        :  *.:.*: ************:*************************************

Rgp     YIPDGVADVTAQKPYTLTVVGKTITVTCQGEAMIYDMNGRRLAAGRNTVVYTAQGGYYAV 1687
Kgp     YIPDGVADVTAQKPYTLTVVGKTITVTCQGEAMIYDMNGRRLAAGRNTVVYTAQGGYYAV 1707
        ************************************************************

Rgp     MVVVDGKSYVEKLAVK 1703
Kgp     MVVVDGKSYVEKLAIK 1723
        **************:*
```

Fig 9: Shows the ClustalW sequence alignment of Rgp and Kgp from the bacteria P. gingivalis.

Both these sequences are conserved for the major part of the sequence right from residue numbers 858-882 to the end of the C- terminal with a limited number of deviations in between, though the differences in the sequences at the N-terminal region can be hypothesized as the reason for their specificity to cleave after an arginine and lysine residue respectively. Further more there are a lot of gaps in the alignment showing the mutations that might have occurred.

Since the gingipains activate proMMPs such as proMMP-9, similar to PgtE of Salmonella, a sequence alignment between the gingipains and PgtE was carried out using ClustalW to find if there were any conserved regions.

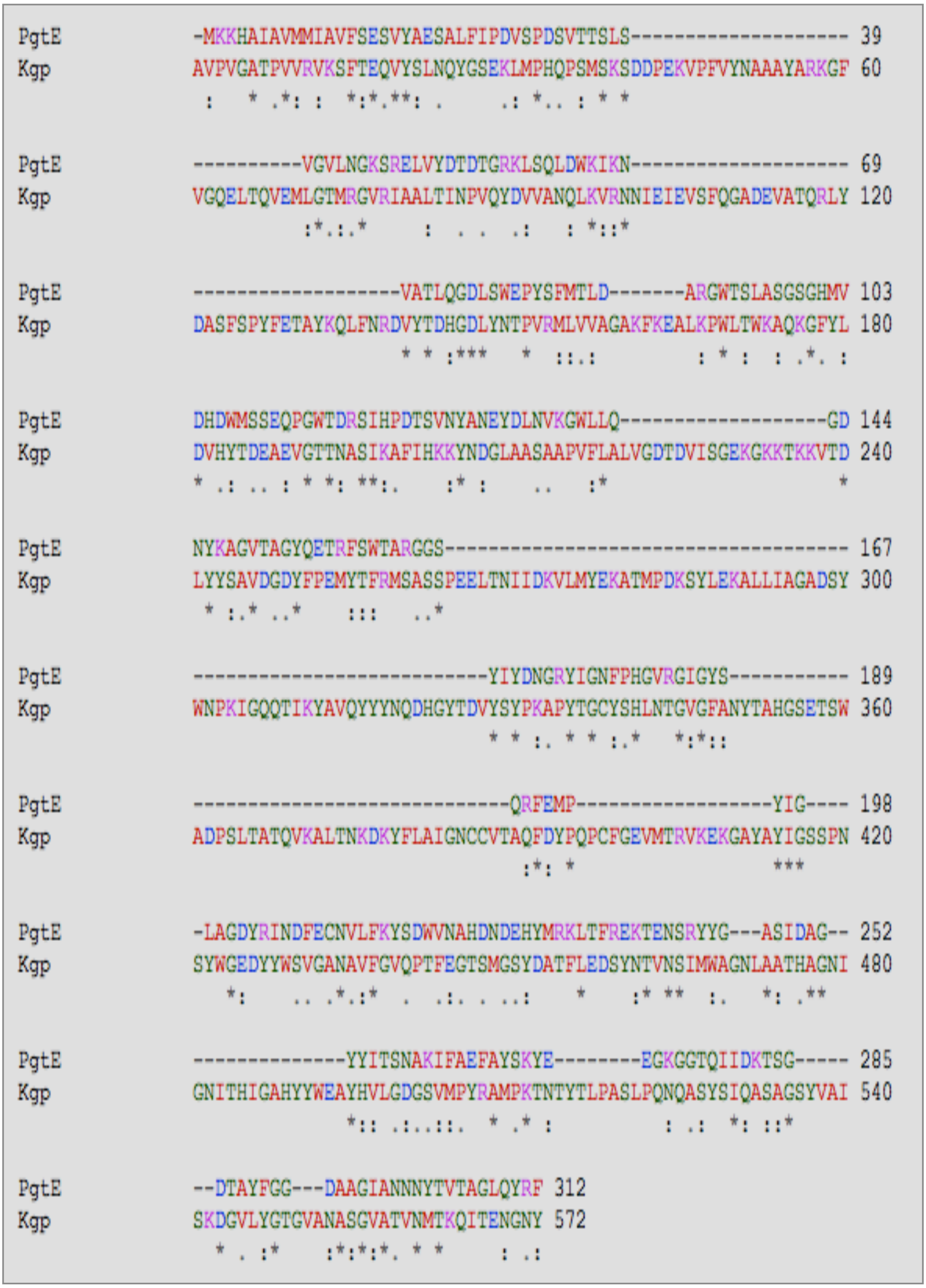

Fig 10: Shows the sequence alignment between Kgp of P. gingivalis and PgtE of Salmonella (only part of the Kgp sequence corresponding to PgtE is shown, the complete Kgp sequence is listed in Appendix 2).

```
PgtE            ------------------MKKHAIAVMMIAVFSESVYAESALFIPDVSPDSVTTSLSVGV 42
RgpA            FVDWKNQRGLRTEVKVAEDIASPVTANAIQQFVKQEYEKEGNDLTYVLLVGDHKDIPAKI 60
                                      .::.  *  * :. * :..  :. *   .  ..:.. :

PgtE            LNGKSRELVYDTDTG------------------RKLSQLDWKIKNVATLQGDLSWEPYSF 84
RgpA            TPGIKSDQVYGQIVGNDHYNEVFIGRFSCESKEDLKTQIDRTIHYERNITTEDKWLGQAL 120
                  * . : **.  .*                     :*:* .*:   .:  : .*   ::

PgtE            MTLDARGWTSLASG---------------------------------------------- 98
RgpA            CIASAEGGPSADNGESDIQHENVIANLLTQYGYTKIIKCYDPGVTPKNIIDAFNGGISLV 180
                   .*.* .*  .*

PgtE            --SGHMVDHDWMSSEQ-------------------------------PGWTDRSIHPDTS 125
RgpA            NYTGHGSETAWGTSHFGTTHVKQLTNSNQLPFIFDVACVNGDFLFSMPCFAEALMRAQKD 240
                  :**  :  * :*.                                * :::  ::.:..

PgtE            VN--------------------------------------------------------YA 129
RgpA            GKPTGTVAIIASTINQSWASPMRGQDEMNEILCEKHPNNIKRTFGGVTMNGMFAMVEKYK 300
                 :                                                        *

PgtE            NEYDLNVKGWLLQGD------------------------NYKAGVTAGYQETRFSWTARG 165
RgpA            KDGEKMLDTWTVFGDPSLLVRTLVPTKMQVTAPAQINLTDASVNVSCDYNGAIATISANG 360
                :: :  :. * : **                        : ...*:..*: :  : :*.*

PgtE            ---GSYIYDNGRYIGNFP-------HGVRGIGYSQRFEMPYIGLAG---------DYRIN 206
RgpA            KMFGSAVVENGTATINLTGLTNESTLTLTVVGYNKETVIKTINTNGEPNPYQPVSNLTAT 420
                   ** : :**    *:.         :  :**.:.  :  *.  *         :   .

PgtE            DFECNVLFKYSDWVNAHDNDEHYMR---------KLTFREKTENSR-------------- 243
RgpA            TQGQKVTLKWDAPSTKTNATTNTARSVDGIRELVLLSVSDAPELLRSGQAEIVLEAHDVW 480
                    :* :*:.   .  :   :  *          *:. : .*  *

PgtE            ---------------YYGASIDAGYYITSN-----AKIFAEFAYSKYEEGK--------- 274
RgpA            NDGSGYQILLDADHDQYGQVIPSDTHTLWPNCSVPANLFAPFEYTVPENADPSCSPTNMI 540
                                **  * :. :         *::** * *:  *:..

PgtE            -GGTQIIDKTSGDTAYFGGDAAG-----------IANNNYTVTAGLQYRF---------- 312
RgpA            MDGTASVNIPAGTYDFAIAAPQANAKIWIAGQGPTKEDDYVFEAGKKYHFLMKKMGSGDG 600
                 .**  :: .:*   :  . . .             :::*.. ** :*:*
```

Fig 11: Showing the sequence alignment between RgpA of P. gingivalis and PgtE of Salmonella (only part of RgpA sequence that is aligned to the PgtE is shown, the complete RgpB sequence is listed in Appendix 3).

```
PgtE            MKKHAIAVMMIAVFS----ESVYAESALFIPDVSPDSVTTSLSVGVLNGKSRELVY-DTD 55
RgpB            MKKNFSRIVSIVAFSSLLGGMAFAQPAERGRNPQVRLLSAEQSMSKVQFRMDNLQFTDVQ 60
                ***:   :: *..**      .:*:.*    : .   :::. *:. :: :  :* : *.:

PgtE            TGRKLSQLDWKIKNVATLQG---DLSWEPYSFMTLDARGWTSLASGSGHMVDHDWMSSEQ 112
RgpB            TSKGVAQVPTFTEGVNISEKGTPILPILSRSLAVSETRAMKVEVVSSKFIEKKDVLIAPS 120
                *.: ::*:    :.*   :     *.  . *: . ::*. .  . .* .: .:* : : .

PgtE            PGWTDRSIHPDTS-VNYANEYDLN------------------------------------ 135
RgpB            KGVISRAENPDQIPYVYGQSYNEDKFFPGEIATLSDPFILRDVRGQVVNFAPLQYNPVTK 180
                 *  .*: :**     *.:.*: :

PgtE            -------------------------VKGWLLQG-DNYKAGVTAGYQETRFSWTAR---GG 166
RgpB            TLRIYTEIVVAVSETAEAGQNTISLVKNSTFTGFEDIYKSVFMNYEATRYTPVEEKENGR 240
                                         **.  : * ::   .*  .*: **:: . .   *

PgtE            SYIYDNGRYIGNFP-----HGVRGIG--------------------------YSQRFEMP 195
RgpB            MIVIVAKKYEGDIKDFVDWKNQRGLRTEVKVAEDIASPVTANAIQQFVKQEYEKEGNDLT 300
                  :    :* *::      :. **:                            .:  ::.

PgtE            YIGLAGDYR-------------------INDFECNVLFK-------YSDWVNAHDNDEHY 229
RgpB            YVLLVGDHKDIPAKITPGIKSDQVYGQIVGNDHYNEVFIGRFSCESKEDLKTQIDRTIHY 360
                *: *.**::                   :.: . * :*         .*  .  *.  **

PgtE            MRKLTFREKTENSRYYGASIDAGYYITSN----------AKIFAEFAYSK----YEEGKG 275
RgpB            ERNITTEDKWLGQALCIASAEGGPSADNGESDIQHENVIANLLTQYGYTKIIKCYDPGVT 420
                 *::* .:*  ..    ** :.*    ..          *::::::.*:*    *: *

PgtE            GTQIIDKTSG------------DTAYFGGDAAGIANNNYTVTAGLQYRF----------- 312
RgpB            PKNIIDAFNGGISLVNYTGHGSETAWGTSHFGTTHVKQLTNSNQLPFIFDVACVNGDFLF 480
                 .:***  .*            :**:  .. .    :: * :  * : *

PgtE            --------------------
RgpB            SMPCFAEALMRAQKDGKPTG 500
```

Fig 12: Showing the sequence alignment between RgpB of P. gingivalis and PgtE of Salmonella (only the part of RgpB sequence that aligns with PgtE is shown, the complete RgpB sequence is listed in Appendix 4)

All three gingipains share a few conserved domains with PgtE of Salmonella, with the lysine specific cysteine protease Kgp getting the highest alignment score, RgpB and RgpA lying after that respectively. Also a lot of gaps can be identified in all three alignments, which might have been mutations during the evolutionary process.

# 4. DISCUSSION

Bacterial proteases are widely spread among all types of bacteria present; they take part in important functions like evasion and colonization of the host's immune system, intake of nutrients for proliferation and growth, aiding in dissemination and causing damage to the tissue at the time of infection (Miyoshi and Shinoda, 1997). They also play other subtle functions, which have been recently identified during the interaction between bacteria and the host organism, from the cleavage of receptors on the cell surface, protease inhibitor inactivation, stopping cytokine networks and disruption of cascade activation pathways (Travis and Potempa, 2000). Our study primarily focused on finding bacterial proteases that could prove to be helpful for that particular process, site directed drug delivery. The study was based on a few recent interesting research results, which identified bacterial proteases with activity that could be utilized in the way proposed. Of the bacterial proteases of interest, three particular proteases were chosen as the primary targets of study, PgtE of Salmonella enterica serovar Typhimurium, PrtA of Photorhabdus luminescens and the Gingipains of Porphyromonas gingivalis. These proteases where chosen particularly because of their specific activity on some host components such as the extracellular matrix (ECM) and other proteases in the host system as well. PgtE for example was studied to have activity on the ECM and more importantly it was able to activate proMMP-9 as well as mimic the activities of MMPs (Ramu *et al.*, 2007). PrtA act on a class of proteins, which have been identified to have immunological activity and the gingipains, RgpA, RgpB and Kgp, also degrade the ECM and activate proMMPs.

Multiple sequence alignment of PgtE using ClustalW, with four different proteases (Pla of Y. pestis, OmpT of E. coli, OMP of E. clocae and OMP of Erwinia sp.) obtained through BLAST search, yielded results that showed a high range of conserved regions among the proteases. Of which Pla and OmpT had higher sequence similarity, which was understandable because they belong to the same family of omptins, which was further proven by phylogenetic analysis. All the proteases that were used to compare with PgtE were observed to have activity on the ECM components. Pla was observed to be good plasminogen activator, which is similar to the function of PgtE activating plasminogen, though the extent of plasminogen activation by Pla was much higher. It should be noted that Pla induces ECM degradation via the activation of plasminogen into plasmin (Lähteenmäki *et al*., 2005). Also the reduced effect of PgtE on plasminogen when compared to Pla, is due to structural adaptations and actually helps Salmonella in recognizing the host macrophage cell produced proMMP-9 (Ramu *et al*., 2007). OmpT has relatively less in plasminogen activation compared to Pla and PgtE. The ability of PgtE to activate proMMP-9 and the inability of Pla, which is similar to it, is due to those structural adaptations and sequential differences. It has been found that by small substitutions in Pla, specifically 2 amino acid substitutions in L5 and 27 amino acids in L1 and L3, it could be made to act as PgtE in the activation of proMMP-9. It has been observed that the loops, L1, L3 and L5 have an important role in the proMMP-9 activation by PgtE (Ramu *et al*., 2007). Structural alignment and analysis also showed the differences in the structure and sequence of Pla and PgtE, which could be purported as the reason for different substrate specifity. The other important activity, which has been unique to it among these proteases studied, is the mimicking of MMPs. PgtE is able mimic the activity of MMP-9 and act as a gelatinase in addition to activating the MMP-9 precursor. Its ability to cleave gelatin has been identified, the loops L1 and L5 playing an important role in it. Another important thing to be noted is that there is a distinct similarity in the loops present in the hemopexin domain of PgtE and MMP-3 (stromelysin-1). It has been already proved that MMP-3 activates proMMP-9, cleaves gelatin and also inactivates the protease inhibitor $\alpha$2-antiplasmin (Lijnen *et al*., 2001). FPHGVRGI of L3 and IIDKTSGDTAYF of L5 in PgtE align to the amino acid sequences FPPTVRKI and ISDKEKNKTYFF of MMP-3.

When designing a possible structural motif for a peptide substrate from the cleavage data of PgtE, which could be used as a prodrug carrier, the above said recommendations can be taken into account. It could be suggested that if the peptide substrate on which the prodrug is planned to be attached should have some general specifications so that it would be cleaved by PgtE. The known substrate specifity of PgtE, as well as, substrate specificities of proteases having similar function like PgtE can be used for that purpose. For example, PgtE has a calcium binding EF- hand domain, which proves that it might bind to the calcium ions if found in its substrates. This can be further proved by the fact that, proMMP-9 which is a substrate for PgtE does have calcium ions in it (Kimata *et al*., 2006). Similarly the preference of PgtE in its substrate for specific prime site residues can also be taken as an example for this.

PrtA of Photorhabdus luminescens is a serralysin that helps in the virulence of the bacteria and is involved in the early infection stages of the host organism (Marokhazi *et al*., 2004). PrtA action on a family of proteins called PAT proteins was analysed, in the entire set of PAT proteins that PrtA cleaves six are known to have immunological functions. ClustalW analysis was carried out between the known PAT proteins to understand any similarity in the sequences. Though they have similar functions the conserved regions among the proteins were very minimal, which does not elucidate any substrate specificity on the part of PrtA. Phylogenetic analysis was also done to study the evolutionary divergence among the PAT proteins. The specifity of the cleavage site of PrtA was studied in hydrolysis of insulin A, B and β-lipotropin. PrtA was observed to show preference to hydrophobic amino acids in specific positions unlike other serralysins, also another important result observed was PrtA could even cleave peptides as short as six amino acids (Marokhazi *et al*., 2007). Further it was evident that PrtA required interaction with substrates on the either side of the scissile bond for efficient cleavage during peptide hydrolysis. In biological peptides, PrtA has higher preference for aliphatic residues and a preference in choosing valine over leucine at the substrate position P2’ which is similar to other serralysins (Louis et al., 1999). These properties and specifications of PrtA should be taken into account when considering developing peptide substrates of PrtA as prodrug carriers.

The gingipains were taken into the study as they were comparable to PgtE in their function, both Rgp and Kgp activated proMMPs. ClustalW was used to align Rgp and Kgp to study the conserved regions, as well as the gingipains were aligned with PgtE individually to identify any conserved regions. The results showed that Kgp was more conserved towards PgtE than Rgp; though they are entirely different proteases the conserved regions found among their sequences could be the reason for their related function. Previous studies have shown the specifity of gingipains in cleaving the substrates having particular prime site residues (Ally *et al*., 2003). Identification of specificity in P2′ and P3′ regions observed during the cleavage of gingipains in it substrates, helps to define the properties of their cleavage sites and predict the modelling of a suitable structural motif.

So by studying the known and other possible substrates of these proteases one could suggest, substrate specificities, cleavage site specifics and important residues in the substrate that are required for cleavage, to define a structural motif that can be used to model novel prodrug carriers.

# 5. CONCLUSION

The importance of bacterial proteases on virulence of bacteria gave rise to a wide study of potential protease inhibitors other than the natural ones present in the immune system. Although the study on harnessing these proteases as tools for drug delivery has not been that widely researched, it promises to be a new and exciting way of combating bacterial diseases. The main goal of this study was to identify and understand a number of bacterial proteases that have specific functions such as degradation of ECM and activation of proMMPs to assess them for the possible use in drug delivery purposes. The primary proteases of study, PgtE of Salmonella enterica and PrtA of Photorhabdus luminescens were analysed using their sequence data, along with extensive study on published literature. Their structural and sequential importance towards the substrate specificity has been addressed; in addition, other related proteases having similar functions were also elaborated on. This project work forms the beginning of an extensive project which would likely involve the creation of peptide libraries and large scale cleavage studies of selected proteases and synthetic protein carriers, with use of recombinant enzymes, to help to create a novel prodrug carrier that can be activated in the microenvironments of cancer or infected human tissues. This study has identified the substrate specificities and cleavage site analysis of a selection of such proteases, which form the foundation for further research.

## Acknowledgements

First of all I would like to thank God and my loving parents who helped me to reach the place where I am now.

I would like to express my sincere thanks to Dr. David Mincher and Dr. Clare Taylor for giving me this wonderful opportunity to do my project on a topic I was really interested in. I would like to also thank them for the amazing support and guidance provided throughout the period of my project. Their constant support and encouragement have given me the spirit to do the project to my best capability

I also want to thank Dr. Agnes Turnbull for her support to me. She was always helpful, assisting me whenever I needed her.

Finally I would like to thank my friends who have been with me throughout my life, in happiness and grief. Their constant motivation and support has helped me a lot during this period.

Thank you all.

## 6. REFERENCES

Ally N, Whisstock JC, Sieprawska-Lupa M, Potempa J, Le Bonniec BF, Travis J, Pike RN (2003). Characterization of the specificity of arginine-specific gingipains from Porphyromonas gingivalis reveals active site differences between different forms of the enzymes. *Biochemistry*., **42**(40):11693-700.

Altschul SF, Gish W, Miller W, Myers EW, Lipman DJ (1990). Basic local alignment search tool. *J Mol Biol* **215** (3): 403–410.

Bae KH, Kim IC, Kim KS, Shin YC & Byun SM (1998) The Leu-3 residue of Serratia marcescens metalloprotease inhibitor is important in inhibitory activity and binding with Serratia marcescens metalloprotease. *Arch Biochem Biophys.,* **352**, 37–43.

Barrett AJ, Rawlings ND, Woessner JF, Jr. (1998). Handbook of proteolytic enzymes (CD-ROM). *London: Academic Press,* Chapters 1–569.

Bode W, Gomis-Ruth FX, Stocker W (1993). Astacins, serralysins, snake venom and matrix metalloproteinases exhibit identical zinc-binding environments (HEXXHXXGXXH and Met-turn) and topologies and should be grouped into a common family, the Fmetzincins_. FEBS *Letters.,* **331**:134 – 40.

Bowen D, Blackburn M, Rocheleau T, Grutzmacher C,Frrench-Constant R.H.

(2000). Secreted proteases from Photorhabdus luminescens: separation of the extracellular proteases from the insecticidal Tc toxin complexes. Insect *Microbiol. Mol. Biol*. **30**:69-74.

Bowen D. J, Rocheleau T.A, Grutzmacher C.K, Meslet L, Valens M, Marble D, Dowling, Ffrench-Constant A.R,and M. A. Blight. (2003). Genetic and biochemical characterization of PrtA an RTX-like metalloprotease from Photorhabdus. *Microbiology,* **149**:1581-1591.

Brooks, P. C. (1996). Role of integrins in angiogenesis. *Eur. J. Cancer*, **32A**: 2423–2429.

Brown P.D, (1997). Clinical studies with matrix metalloproteinase inhibitors. *APMIS*., **107**:174–80.

Casarini D.E, Boim M.A, Stella R.C, Schor N, (1999). Endopeptidases (kininases) are able to hydrolyze kinins in tubular fluid along the rat nephron. *Am J Physiol*., **277**: F66–F74.

Castaneda, F.E., Walia, B., Vijay-Kumar, M., Patel, N.R., Roser, S., Kolachala, V.L., Rojas, M., Wang, L., Oprea, G., Garg, P., Gewirtz, A.T., Roman, J., Merlin, D. and Sitaraman, S.V. (2005) Targeted deletion of metalloproteinase 9 attenuates experimental colitis in mice: central role of epithelial-derived MMP. *Gastroenterol.,* **129**, 1991-2008.

Chambers, A. F. and Matrisian L. M. Changing views of the role of matrix metalloproteinases in metastasis (1997). *J. Natl. Cancer Inst. (Bethesda),* 89:1260 –1270.

Chau Y, Dang N.M, Tan F.E, Langer R. (2006). Investigation of targeting mechanism of new dextran-peptide-methotrexate conjugates using biodistribution study in matrix-metalloproteinase over expressing tumor xenograft model. *J. Pharm. Sci.,* 95 (3), 542–551.

Chau Y, Padera R. F, Dang N. M, Langer R (2006). Antitumor efficacy of a novel polymer-peptide-drug conjugate in human tumor xenograft models. *Int. J. Cancer*, 118 (6), 1519– 1526.

Chau Y, Tan F.E, Langer R (2004). Synthesis and Characterization of Dextran-Peptide-Methotrexate Conjugates for Tumor Targeting via Mediation by Matrix Metalloproteinase II and Matrix Metalloproteinase IX. *Bioconjugate Chem*. 15 (4), 931– 941.

Chen H and Schifferli M, (2003). Construction, characterization and immunogenicity of an attenuated Salmonella enterica serovar Typhimurium pgtE vaccine expressing fimbriae with integrated viral epitopes from the spiC promoter. *Infect. Immun.* 71: 4664-4673.

Chenna R, Sugawara H, Koike T, Lopez R, Gibson TJ, Higgins DG, Thompson JD (2003). Multiple sequence alignment with the Clustal series of programs. *Nucleic Acids Res* **31** (13): 3497–3500.

Coussens L.M, Fingleton B, Matrisian L.M (2002). Matrix metalloproteinase inhibitors and cancer: trials and tribulations. *Science*. 295: 2387–2392

Crystal R. G (1990). Alpha-1-antitrypsin deficiency, emphysema, and liver disease: genetic basis and strategies for therapy. *J. Clin. Invest.* 85: 1343-1352.

Daborn J. D, Waterfield N, Blight M.A and Ffrench-Constant R.H (2001). Measuring virulence factor expression by the pathogenic bacterium Photorhabdus luminescens in culture and during insect infection. *J. Bacteriol*., 183:5834-5839.

DeCarlo A. A, Windsor L. J, Bodden M. K, Harber G. J, Birkedal- Hansen B and Birkedal-Hansen H, (1997). Activation and novel processing of matrix metalloproteinases by a thiol-proteinase from the oral anerobe *Porphyromonas gingivalis. J. Dent. Res*. 76:1260–1270.

DeClerck Y.A, Mercurio A.M, Stack MS, Chapman H.A, Zutter M.M, Muschel R.J, Raz A, Matrisian L.M, Sloane B.F, Noel A, Hendrix M.J, Coussens L, Padarathsingh M (2004). Proteases, Extracellular Matrix, and Cancer. *Am J Pathol*., 164:1131–1139

Dekker N, Cox R.C, Kramer A and Egmond M.R (2001). Substrate specificity of the integral membrane protease OmpT determined by spatially addressed peptide libraries. *Biochemistry*, 40:1694-1701.

DeMichele M. A. A, Moon D. G, Fenton J. W and Minnear F. L, (1990). Thrombin's enzymatic activity increases permeability of endothelial cell monolayers. *J. Appl. Physiol*. 69:1599–1606.

Devy L, Blacher S, Grignet-Debrus C, Bajou K, Masson V, Gerard R.D, Gils A, Carmeliet G, Carmeliet P, Declerck P.J, Noel A, Foidart J.M (2002). The pro- or anti-angiogenic effect of plasminogen activator inhibitor 1 is dose dependent. *EMBO J*, 16:147–154.

Duchaud E, Rusniok C, Frangeul L, Buchrieser C, Givaudan A, Taourit S, Bocs S, Boursaux-Eude C, Chandler M, Charles JF, Dassa E, Derose R, Derzelle S, Freyssinet G, Gaudriault S, Médigue C, Lanois A, Powell K, Siguier P, Vincent R, Wingate V, Zouine M, Glaser P, Boemare N, Danchin A, Kunst F (2003). The genome sequence of the entomopathogenic bacterium Photorhabdus luminescens. *Nat Biotechnol.* 21(11): 1307-13.

Egeblad M, Werb Z (2002). New functions for the matrix metalloproteinases in cancer progression. *Nat Rev Cancer*, 2:161–174.

Ehrlich P, Bolduan C. (1906). Collected studies on immunity. New York, NY: John Wiley & Sons, 2 (iii–xi) 586.

Elkington P.T.G, O'Kane C.M.and Friedland J.S. (2005) The paradox of matrix metalloproteinases in infectious disease. *Clin. Exp. Immunol.* 142, 12-20

Eliceiri B. P, and Cheresh D. A. (2001) Adhesion events in angiogenesis . *Curr. Opin. Cell Biol.,* 13: 563–568.

Eppinger M, Worsham PL, Nikolich MP, Riley DR, Sebastian Y, Mou S, Achtman M, Lindler LE, Ravel J (2010). Genome sequence of the deep-rooted Yersinia pestis strain Angola reveals new insights into the evolution and pangenome of the plague bacterium. *J Bacteriol.,* 192(6):1685-99.

Fingleton B, Vargo-Gogola T, Crawford H.C, Matrisian L.M. (2001) Matrilysin [MMP-7] expression selects for cells with reduced sensitivity to apoptosis. *Neoplasia,* 3:459–468

Fischer-Le Saux M, Viallard V, Brunel B, Normand P, Boemare N.E, (1999). Polyphasic classification of the genus Photorhabdus and proposal of new taxa: P. luminescens subsp. luminescens subsp. nov., P. luminescens subsp. akhurstii subsp. nov., P. luminescens subsp. laumondii subsp. nov., P. temperata sp. nov., P. temperata subsp. temperata subsp. nov. and P. asymbiotica sp. nov. *Int J Syst Bacteriol*., 49:1645–56.

Fitzpatrick R.E, Wijeyewickrema L.C, Pike R.N (2009). The gingipains: scissors and glue of the periodontal pathogen, Porphyromonas gingivalis. Future Microbiol., 4(4):471-87.

Fox C.H (1992). New considerations in the prevalence of periodontal disease. *Curr Opin Dent.* 2:5–11.

Fox M and Walsh E (2008). Report sees 7.6 million global 2007-cancer deaths. *Reuters.,* http://www.reuters.com/article/idUSN1633064920071217

Franks, L. M, Teich N. M. (1997), Introduction to the Cellular and Molecular Biology of Cancer, Oxford University Press.

Fridman R, (2006). Metalloproteinases and Cancer. *Cancer Metastasis Rev*., 25: 7–8

Gerweck L. E. Seetharaman K (1996). Cellular pH gradient in tumor versus normal tissue: Potential exploitation for the treatment of cancer. *Cancer Res*., 56 (6), 1194–1198

Giammanco G.M, Pignato S, Mammina C, Grimont F, Grimont P.A, Nastasi A, Giammanco G (2002). Persistent endemicity of *Salmonella bongori* 48:z(35):-- in Southern Italy: molecular characterization of human, animal, and environmental isolates. *J Clin Microbiol*., 40(9):3502-5.

Ginalski K. (2006). Comparative modeling for protein structure prediction. *Curr Opin Struct Biol* **16**(2): 172-7.

Guina T, Yi E.C, Wang H, Hackett M, Miller S.I, (2000). A PhoP-regulated outer membrane protease of Salmonella enterica serovar typhimurium promotes resistance to alpha-helical antimicrobial peptides. *J Bacteriol.*, 182(14):4077-86.

Hanahan D, Weinberg R.A, (2000) The hallmarks of cancer. *Cell*, 100: 57–70

Handley, S.A. and Miller, V.L. (2007) General and specific host responses to bacterial infection in Peyer's patches: a role for stromelysin-1 (matrix metalloproteinase-3) during Salmonella enterica infection. *Mol. Microbiol*. 64, 94-110.

Harper E, Bloch K.J, Gross J (2000). The zymogen of tadpole collagenase. *Biochemistry*, 10:3035–3041.

Hashimoto Y, (1983). Protein, Nucleic Acid and Enzyme, 28, 1220.

Hatakeyama H, Akita H, Kogure K, Oishi, M, Nagasaki Y, Kihira Y, Ueno M, Kobayashi H, Kikuchi H, Harashima H (2007). Development of a novel systemic gene delivery system for cancer therapy with a tumor-specific cleavable PEG-lipid. *Gene Ther.* 14 (1), 68–77.

Imamura T, Potempa J, Pike R. N, Moore J.N, Barton M.H, Travis J. (1995). Effect of free and vesicle-bound cysteine proteinases of Por- phyromonas gingivalis on plasma clot formation: implications for bleeding tendency at periodontitis sites. *J. Biol. Chem.,* 274:18984– 18991.

Imamura, T, Banbula A, Pereira P.J.B, Travis J, Potempa J. (2001). Activation of human prothrombin by arginine-specific cysteine proteinases (gingipains R) from Porphyromonas gingivalis. *J. Biol. Chem.,* 275:18984– 18991.

Johansson N, Ahonen M and Kähäri V.M (2000) Matrix metalloproteinases in tumor invasion. *Cell. Mol. Life. Sci.,* 57, 5-15.

Jones C.B, Sane D.C, Herrington D.M, (2003). Matrix metalloproteinases: a review of their structure and role in acute coronary syndrome. *Cardiovasc Res*., 59(4): 812-23.

Kang K, Chung J.H, Kim J (2009). Evolutionary Conserved Motif Finder (ECMFinder) for genome-wide identification of clustered YY1- and CTCF-binding sites. *NucleicAcids Res.* 37(6): 2003-13

Kimata M, Ishizaki M, Tanaka H, Nagai H, Inagaki N (2006). Production of matrix metalloproteinases in human cultured mast cells: involvement of protein kinase C-mitogen activated protein kinase extracellular signal-regulated kinase pathway. *Allergol Int*., 55(1): 67-76

Knodler L.A. and Finlay B.B. (2001) Salmonella and apoptosis: to live or let die? *Microbes infect.,* 3, 1321- 1326.

Koblinski J.E, Ahram M, Sloane B.F, (2000). Unraveling the role of proteases in cancer. *Clin Chim Acta*., 291(2):113-35.

Kukkonen M and Korhonen T.K. (2004) The omptin family of enterobacterial surface proteases/adhesins: from housekeeping in Escherichia coli to systemic spread of Yersinia pestis*. Int. J. Med. Microbiol.,* 294, 7-14.

Kukkonen M, Lähteenmäki K, Suomalainen M, Kalkkinen N, Emödy L, Lång H and Korhonen T.K. (2001) Protein regions important for plasminogen activation and inactivation alpha2-antiplasmin in the surface protease Pla of Yersinia pestis. *Mol. Microbiol.* 40, 1097-1111.

Kramer R. A, and Trail P. A (2001). Inhibition of angiogenesis and metastasis in two murine models by the matrix metalloproteinase inhibitor. *Cancer Res.,* 61: 8480–8485.

Lantz M.S, (1997). Are bacterial proteases important virulence factors? *J Periodontal Res.,* 32(1.2): 126-32.

Lee S, Park K, Lee S.Y, Ryu J. H, Park J. W, Ahn, H. J, Kwon I. C, Youn I.C, Kim K, Choi K (2008). Dark Quenched Matrix Metalloproteinase Fluorogenic

Probe for Imaging Osteoarthritis Development in Vivo. *Bioconjugate Chem.,* **19** (9), 1743–1747.

Leiros H.K, Brandsdal B.O, Andersen O.A, Os V, Leiros I, Helland R, Otlewski J, Willassen NP, Smalås A.O, (2004). Trypsin specificity as elucidated by LIE calculations, X-ray structures, and association constant measurements. *Protein Sci*., **13** (4): 1056–70.

Liotta L.A, Kohn E.C, (2001). The microenvironment of the tumour-host interface. *Nature*, **411**:375–379.

Lijnen, H.R., Van Hoef, B. and Collen, D. (2001) Inactivation of the serpin $\alpha$2-antiplasmin by stromelysin-1. *Biochem. Biophys. Acta*. **1547**, 206-213.

Louis D, Bernillon J & Wallach J.M, (1999) Use of a 49-peptide library for a qualitative and quantitative determination of pseudomonal serralysin specificity. *Int J Biochem Cell Biol* **31**, 1435–1441.

Maeda H (1996). Role of microbial proteases in pathogenesis. *Microbiol Immunol*., **40**:685–699.

Mahmood U, Weissleder R (2003). Near-infrared optical imaging of proteases in cancer. *Mol Cancer Ther.*, **2**(5):489-96.

Marokházi, J., K. Lengyel, S. Pekár, G. Felfoldi, A. Patthy, L. Gráf, A. Fodor,I.

Venekei. (2004). Comparison of proteolytic activities produced by entomopathogenic Photorhabdus bacteria: strain- and phase-dependent heterogeneity in composition and activity of four enzymes. *Appl. Environ. Micro- biol.* 70:7311–7320.

Marti-Renom M.A, Stuart A.C, Fiser A, Sanchez R, Melo F, Sali A. (2000). Comparative protein structure modelling of genes and genomes. *Annu Rev Biophys Biomol Struct* 29: 291-325.

McCawley L.J and Matrisian L.M. (2001) Matrix metalloproteinases: they're not just for matrix anymore. *Curr. Opin. Cell Biol*., 13, 534-540.

McClelland M, Sanderson K.E, Spleth J, Clifton S.W, Latreille P, Courtney L, Porwollik, S, Ali J, Dante M, Du F, Hou S, Layman D, Leonard S, Nguyen, C., Scott K, Holmes A, Grewal N, Mulvaney E, Ryan E, Sun H, Florea L, Miller W, Stoneking T, Nhan M, Waterston R and Wilson R.K. (2001) Complete genome sequence of Salmonella enterica serovar Typhimurium LT2. *Nature,* 413, 852-856.

Miyoshi S.I, Shinoda S (1997). Bacterial metalloproteases as the toxic factor in infection. *J Toxicol Toxin Rev*.16: 177–194.

Mok H, Bae K. H, Ahn C.H, Park T. G. (2009) PEGylated and MMP-2 Specifically DePEGylated Quantum Dots: Comparative Evaluation of Cellular Uptake. *Langmuir*., 25 (3), 1645– 1650.

Mount D.M. (2004). *Bioinformatics: Sequence and Genome Analysis* (2nd ed.)

Nagase H, Visse R, Murphy G (2006). Structure and function of matrix metalloproteinases and TIMPs. *Cardiovasc Res.*, 69(3): 562-73.

Nagase H, Woessner JF (1999). "Matrix metalloproteinases." *J. Biol. Chem.* 274 (31): 21491–4.

Naglich J. G, Jure-Kunkel M, Gupta E, Fargnoli J, Henderson A. J, Lewin A. C, Talbott R, Baxter A, Bird J, Savopoulos R, Wills R, Noe V, Fingleton B, Jacobs K, Crawford HC, Vermeulen S, Steelant W Bruyneel E, Matrisian LM, Mareel M (2001). Release of an invasion promoter E-cadherin fragment by matrilysin and stromelysin-1. *J Cell Sci.,* 114:111–118

Naito M, Hirakawa H, Yamashita A, Ohara N, Shoji M, Yukitake H, Nakayama K, Toh H, Yoshimura F, Kuhara S, Hattori M, Hayashi T, Nakayama K (2008). Determination of the genome sequence of Porphyromonas gingivalis strain ATCC 33277 and genomic comparison with strain W83 revealed extensive genome rearrangements in P. gingivalis. *DNA Res*.,15(4):215-25.

O'Brien-Simpson N M, Black C L, Bhogal P S, Cleal S M, Slakeski N, Higgins T J, Reynolds E C (2000). Serum IgG and IgG subclass responses to the RgpA-Kgp proteinase-adhesin complex of P. gingivalis in adult periodontitis. *Infect Immun.,* 68:2704–2712

Ochman H. and Wilson A.C. (1987) Evolution in bacteria: evidence for a universal substitution rate in cellular genomes. *J. Mol. Evol.,* 26, 74-86.

Okamoto T, Akaike T, Suga M, Tanase S, Horie H, Miyajima S, Ando M, Ichinose Y and Maeda H. (1997) Activation of human matrix metalloproteinases by various bacterial proteinases. *J. Biol. Chem.* 272, 6059-6066.

Onder O, Turkarslan S, Sun D, Daldal F (2008). Overproduction or absence of the periplasmic protease DegP severely compromises bacterial growth in the absence of the dithiol: disulfide oxidoreductase DsbA. *Mol Cell Proteomics*. 7(5):875-90.

Opdenakker G, Van den Steen P.E, Dubois B, Nelissen I, Van Coillie E, Masure S, Proost P and Van Damme J. (2001) Gelatinase B functions as regulator and effector in leukocyte biology*. J. Leukoc. Biol.* 69, 851-859.

Overall C.M. (2002). Molecular determinants of metalloproteinase substrate specificity—matrix metalloproteinase substrate bind- ing domains, modules, and exosites. *Mol Biotechnol*., 22:51–86.

Overall C. M and Lo´pez-Otı´n C (2002). Strategies for MMP inhibition in cancer: Innovations for the post-trial era. *Nat. Rev. Cancer.,* 2: 657–672.

Pattamapun K, Tiranathanagul S, Yongchaitrakul T, Kuwatanasuchat J and Pavasant P. (2003) Activation of MMP-2 by Porphyromonas gingivalis in human periodontal ligament cells. *J. Periodontal Res.,* 38, 115-121.

Popadiak K, Potempa J, Riesbeck K, Blom AM (2007). Biphasic effect of gingipains from Porphyromonas gingivalis on the human complement system. *J Immunol*., 178(11):7242-50.

Revazishvili T, Rajanna C, Bakanidze L, Tsertsvadze N, Imnadze P, O'Connell K, Kreger A, Stine O.C, Morris J.G Jr, Sulakvelidze A (2008). Characterisation of Yersinia pestis isolates from natural foci of plague in the Republic of Georgia, and their relationship to *Y. pestis* isolates from other countries. *Clin Microbiol Infect.*14 (5): 429-36.

Rice S.A, Givskov M, Steinberg P, Kjelleberg S (1999). Bacterial signals and antagonists: the interaction between bacteria and higher organisms. *J Mol Microbiol Biotechnol.,*1:23–31.

Richter-Dahlfors A, Buchan A.M.J, Finlay B.B. (1997). Murine salmonellosis studied by confocal microscopy: Salmonella typhimurium resides intracellularly inside macrophages and exerts a cytotoxic effect on phagocytes in vivo. *J. Exp. Med.* 186:569.

Savagner P (2001). Leaving the neighbourhood: molecular mechanisms in-volved during epithelial-mesenchymal transition. *Bioessays*, 23: 912–923

Serkina A.V, Shevelev A.B, Chestukhina G.G, (2001). Structure and functions of bacterial proteinase precursors. *Bioorg Khim*. 27(5): 323-46.

Skyberg J.A, Johnson T.J, Johnson J.R, Clabots C, Logue C.M, Nolan L.K (2006). Acquisition of avian pathogenic Escherichia coli plasmids by a commensal E. coli isolate enhances its abilities to kill chicken embryos, grow in human urine, and colonize the murine kidney. *Infect Immun.,* 74(11): 6287-92.

Sodeinde O.A. and Goguen J.D. (1989) Nucleotide sequence of the plasminogen activator gene of Yersinia pestis: relationship to ompT of Escherichia coli and gene E of Salmonella typhimurium. *Infect. Immun.,* 57, 1517-1523.

Stetler-Stevenson W. G (1999). Matrix metalloproteinases in angiogenesis: a moving target for therapeutic intervention. *J. Clin. Investig.,* 103: 1237–1241.

Stormo G. D (2000). "DNA binding sites: representation and discovery". *Bioinformatics* **16** (1): 16–23.

Suomalainen M, Haiko J, Ramu P, Lobo L, Kukkonen M, Westerlund-Wikström B, Virkola R, Lähteenmäki K, Korhonen TK (2007). Using every trick in the book: the Pla surface protease of Yersinia pestis. *Adv Exp Med Biol*. 603:268-78

Tam E.M, Morrison C.J, Wu YI, Stack M.S, Overall C.M. (2004).Membrane protease proteomics: Isotope-coded affinity tag ms identification of undescribed mt1-matrix metalloproteinase substrates. *Proc Natl Acad Sci*., 101:6917–6922.

Thompson J.D., Higgins D.G. and Gibson T.J. (1994) CLUSTAL W: improving the sensitivity of progressive multiple sequence alignments through sequence weighting, position specific gap penalties and weight matrix choice. *Nucl. Acids Res.,* 22:4673-4680.

Tokuda M, Karunakaran T, Hamada N & Kuramitsu H. (1998) Role of Arg-gingipain A in virulence of Porphyromonas gingivalis. *Infection and Immunity*, 66, 1159–1166.

Travis J, Potempa J (2000). Bacterial proteinases as targets for the development of second-generation antibiotics. *Biochem Biophys Acta.,* 1477:35–50.

Travis J, Pike R, Imamura T, Potempa J (1997). Porphyromonas gingivalis proteinases as virulence factors in the development of periodontitis. *J Periodont Res*., 32:120–125.

Tsiodras S, Mantzoros C, Hammer S, Samore M (2000). Effects of protease inhibitors on hyperglycemia, hyperlipidemia, and lipodystrophy: a 5-year cohort study. *ArchIntern Med.,* 160(13):2050-6.

Vandeputte-Rutten L, Kramer R.A, Kroon J, Dekker N, Egmond M.R. and Gros P. (2001) Crystal structure of the outer membrane protease OmpT from Escherichia coli suggests a novel catalytic site. *EMBO J.* **20**, 5033-5039.

Vartak D.G, Gemeinhart R.A, (2007). Matrix mettaloproteases: underutilized targets for drug delivery. *J Drug Target.,* **15**(1):1-20.

Wang L, Jiang T. (1994) On the complexity of multiple sequence alignment. *J Comput Biol* **1**:337-348.

Watanabe K. (2004) Collagenolytic proteases from bacteria. *Appl. Microbiol. Biotechnol*., **63**, 520-526.

Weissleder R, Tung C. H, Mahmood U, Bogdanov A Jr (1999). In vivo imaging of tumors with protease-activated near-infrared fluorescent probes. *Nat. Biotechnol.,* **17** (4), 375–378.

Werb Z (1997). ECM and cell surface proteolysis: Regulating cellular ecology. *Cell.,* **91** (4), 439-442.

Wimley, W.C., (2003) The versatile β-barrel membrane protein. *Curr. Opin. Struct. Biol.* **13,** 404-411.

Wunder, A., Tung, CH., Muller-Ladner, U., Weissleder, R., Mahmood, U., (2004). In vivo imaging of protease activity in arthritis: a novel approach for monitoring treatment response. *Arthritis Rheum.*, 50 (8), 2459–2465.

Xie, B., Dong, Z., Fidler, I.J.,(1994) Regulatory mechanisms for the expression of type IV collagenases/gelatinases in murine macrophages. *J. Immunol.* 152, 3637-3644.

Zhang Y and Skolnick J. (2005). The protein structure prediction problem could be solved using the current PDB library. *Proc. Natl. Acad. Sci.* USA **102**(4): 1029-34.

# APPENDIX

1. BLASTp Result For PgtE

BLASTP RESULTS

| PgtE | Aligned Seq. | QueryAlign % | ResultAlign% | E value | Score |
|---|---|---|---|---|---|
| gi\|9230677\|gt | gi\|16765721\|r | 100 | 100 | 0 | 646 |
| gi\|9230677\|gt | gi\|168242535 | 99.68 | 99.68 | 0 | 645 |
| gi\|9230677\|gt | gi\|62180966\|r | 99.68 | 99.68 | 0 | 645 |
| gi\|9230677\|gt | gi\|197249873 | 99.36 | 99.36 | 0 | 644 |
| gi\|9230677\|gt | gi\|168817864 | 99.04 | 99.36 | 0 | 642 |
| gi\|9230677\|gt | gi\|168465889 | 99.04 | 99.36 | 0 | 642 |
| gi\|9230677\|gt | gi\|205353509 | 99.04 | 99.36 | 0 | 642 |
| gi\|9230677\|gt | gi\|204928945 | 98.72 | 99.36 | 0 | 640 |
| gi\|9230677\|gt | gi\|16761320\|r | 98.72 | 99.36 | 0 | 640 |
| gi\|9230677\|gt | gi\|168261577 | 98.72 | 99.36 | 0 | 640 |
| gi\|9230677\|gt | gi\|161612865 | 98.4 | 99.36 | 0 | 639 |
| gi\|9230677\|gt | gi\|56412706\|r | 98.4 | 99.36 | 0 | 639 |
| gi\|9230677\|gt | gi\|168237423 | 98.4 | 98.72 | 0 | 639 |
| gi\|9230677\|gt | gi\|160863800 | 89.1 | 94.23 | 5.00E-163 | 577 |
| gi\|9230677\|gt | gi\|228879508 | 89.1 | 94.23 | 5.00E-163 | 577 |
| gi\|9230677\|gt | gi\|154256\|gb\| | 97.83 | 98.56 | 8.00E-160 | 566 |
| gi\|9230677\|gt | gi\|296102467 | 74.36 | 84.62 | 7.00E-137 | 490 |
| gi\|9230677\|gt | gi\|45478719\|r | 72.76 | 83.65 | 8.00E-135 | 484 |
| gi\|9230677\|gt | gi\|146311554 | 72.76 | 84.29 | 8.00E-135 | 483 |
| gi\|9230677\|gt | gi\|258635911 | 72.12 | 82.69 | 2.00E-134 | 482 |
| gi\|9230677\|gt | gi\|16082686\|r | 72.44 | 83.33 | 3.00E-134 | 482 |
| gi\|9230677\|gt | gi\|155525\|gb\| | 72.12 | 83.33 | 3.00E-134 | 481 |
| gi\|9230677\|gt | gi\|206575691 | 73.08 | 82.69 | 9.00E-134 | 480 |
| gi\|9230677\|gt | gi\|167471087 | 72.12 | 83.01 | 2.00E-133 | 479 |
| gi\|9230677\|gt | gi\|213416633 | 99.13 | 99.57 | 8.00E-133 | 477 |
| gi\|9230677\|gt | gi\|213163380 | 99.12 | 99.56 | 2.00E-131 | 473 |
| gi\|9230677\|gt | gi\|31790978\|r | 69.55 | 79.81 | 6.00E-128 | 461 |
| gi\|9230677\|gt | gi\|259910350 | 68.91 | 79.17 | 4.00E-127 | 458 |
| gi\|9230677\|gt | gi\|165940405 | 71.99 | 83.33 | 1.00E-120 | 436 |
| gi\|9230677\|gt | gi\|289812336 | 99.04 | 99.52 | 4.00E-119 | 431 |
| gi\|9230677\|gt | gi\|27228688\|r | 67.95 | 76.92 | 3.00E-118 | 429 |
| gi\|9230677\|gt | gi\|156936553 | 64.58 | 80.81 | 5.00E-105 | 385 |
| gi\|9230677\|gt | gi\|213609086 | 98.33 | 98.89 | 2.00E-100 | 370 |
| gi\|9230677\|gt | gi\|281177706 | 47.19 | 66.56 | 8.00E-80 | 301 |
| gi\|9230677\|gt | gi\|291281507 | 46.88 | 66.25 | 2.00E-79 | 300 |
| gi\|9230677\|gt | gi\|110640796 | 46.88 | 66.25 | 2.00E-79 | 300 |
| gi\|9230677\|gt | gi\|193063338 | 46.88 | 66.25 | 2.00E-79 | 300 |
| gi\|9230677\|gt | gi\|256023823 | 46.88 | 66.25 | 2.00E-79 | 299 |
| gi\|9230677\|gt | gi\|15801391\|r | 46.88 | 65.94 | 3.00E-79 | 299 |
| gi\|9230677\|gt | gi\|16128548\|r | 46.56 | 66.56 | 3.00E-79 | 299 |
| gi\|9230677\|gt | gi\|218557503 | 46.88 | 66.25 | 4.00E-79 | 298 |
| gi\|9230677\|gt | gi\|218703881 | 46.56 | 66.25 | 4.00E-79 | 298 |
| gi\|9230677\|gt | gi\|260866709 | 46.56 | 66.25 | 5.00E-79 | 298 |
| gi\|9230677\|gt | gi\|227884456 | 46.88 | 66.25 | 5.00E-79 | 298 |
| gi\|9230677\|gt | gi\|91209611\|r | 46.56 | 66.25 | 6.00E-79 | 298 |
| gi\|9230677\|gt | gi\|215485608 | 46.56 | 65.94 | 1.00E-78 | 297 |
| gi\|9230677\|gt | gi\|26246544\|r | 46.56 | 65.94 | 3.00E-78 | 296 |
| gi\|9230677\|gt | gi\|148360161 | 46.23 | 66.23 | 1.00E-77 | 294 |
| gi\|9230677\|gt | gi\|54298388\|r | 45.9 | 66.23 | 1.00E-77 | 293 |
| gi\|9230677\|gt | gi\|15988003\|r | 50.17 | 69.07 | 2.00E-77 | 293 |
| gi\|9230677\|gt | gi\|54295228\|r | 45.9 | 65.57 | 2.00E-77 | 293 |
| gi\|9230677\|gt | gi\|270160177 | 45.9 | 64.59 | 2.00E-77 | 293 |

BLASTP RESULTS

| | | | | | |
|---|---|---|---|---|---|
| gi\|9230677\|gl | gi\|170768229 | 45.91 | 66.35 | 7.00E-77 | 291 |
| gi\|9230677\|gl | gi\|52842596\|r | 45.25 | 66.23 | 9.00E-77 | 291 |
| gi\|9230677\|gl | gi\|162317578 | 44.83 | 65.83 | 2.00E-76 | 290 |
| gi\|9230677\|gl | gi\|115512771 | 44.83 | 65.83 | 2.00E-76 | 290 |
| gi\|9230677\|gl | gi\|9507742\|re | 46.56 | 65.94 | 2.00E-76 | 289 |
| gi\|9230677\|gl | gi\|213622007 | 99.28 | 100 | 4.00E-76 | 288 |
| gi\|9230677\|gl | gi\|50983081\|( | 48.62 | 68.62 | 5.00E-76 | 288 |
| gi\|9230677\|gl | gi\|194306026 | 49.65 | 68.53 | 1.00E-74 | 284 |
| gi\|9230677\|gl | gi\|238753031 | 48.59 | 65.49 | 1.00E-72 | 277 |
| gi\|9230677\|gl | gi\|161503662 | 43.71 | 65.41 | 7.00E-72 | 275 |
| gi\|9230677\|gl | gi\|283785705 | 43.44 | 65.62 | 3.00E-71 | 272 |
| gi\|9230677\|gl | gi\|238794252 | 45.66 | 65.27 | 8.00E-70 | 268 |
| gi\|9230677\|gl | gi\|22126834\|r | 42.68 | 63.38 | 1.00E-69 | 267 |
| gi\|9230677\|gl | gi\|186894675 | 42.68 | 63.06 | 3.00E-69 | 266 |
| gi\|9230677\|gl | gi\|170025055 | 42.36 | 62.74 | 2.00E-68 | 263 |
| gi\|9230677\|gl | gi\|51595610\|r | 42.36 | 63.06 | 3.00E-68 | 263 |
| gi\|9230677\|gl | gi\|84060870\|r | 46.05 | 64.95 | 8.00E-68 | 261 |
| gi\|9230677\|gl | gi\|259906940 | 42.41 | 60.44 | 2.00E-67 | 259 |
| gi\|9230677\|gl | gi\|292486737 | 43.24 | 60.47 | 2.00E-63 | 246 |
| gi\|9230677\|gl | gi\|254497640 | 39.27 | 63.04 | 7.00E-62 | 241 |
| gi\|9230677\|gl | gi\|253800991 | 47.58 | 67.34 | 6.00E-60 | 235 |
| gi\|9230677\|gl | gi\|260778787 | 42.04 | 60.51 | 6.00E-60 | 235 |
| gi\|9230677\|gl | gi\|289812096 | 98.23 | 99.12 | 1.00E-58 | 231 |
| gi\|9230677\|gl | gi\|32306984\|( | 38.05 | 59.12 | 7.00E-58 | 228 |
| gi\|9230677\|gl | gi\|187734391 | 38.05 | 59.12 | 8.00E-58 | 228 |
| gi\|9230677\|gl | gi\|81244028\|( | 38.05 | 59.12 | 1.00E-57 | 228 |
| gi\|9230677\|gl | gi\|161986648 | 38.05 | 59.12 | 1.00E-57 | 227 |
| gi\|9230677\|gl | gi\|82524812\|r | 38.05 | 59.12 | 1.00E-57 | 227 |
| gi\|9230677\|gl | gi\|296100145 | 38.36 | 59.75 | 2.00E-57 | 226 |
| gi\|9230677\|gl | gi\|56404036\|r | 38.36 | 59.75 | 2.00E-57 | 226 |
| gi\|9230677\|gl | gi\|194435390 | 37.74 | 58.81 | 5.00E-57 | 225 |
| gi\|9230677\|gl | gi\|15485223\|( | 38.05 | 59.43 | 6.00E-57 | 225 |
| gi\|9230677\|gl | gi\|74315076\|r | 37.74 | 58.81 | 8.00E-57 | 224 |
| gi\|9230677\|gl | gi\|213022460 | 98.99 | 98.99 | 4.00E-52 | 209 |
| gi\|9230677\|gl | gi\|59711853\|r | 39.49 | 55.1 | 2.00E-50 | 203 |
| gi\|9230677\|gl | gi\|292898463 | 34.89 | 52.65 | 6.00E-42 | 175 |
| gi\|9230677\|gl | gi\|292489312 | 34.89 | 52.65 | 9.00E-42 | 174 |
| gi\|9230677\|gl | gi\|292898461 | 34.7 | 53.94 | 9.00E-42 | 174 |
| gi\|9230677\|gl | gi\|292489316 | 34.48 | 52.98 | 9.00E-42 | 174 |
| gi\|9230677\|gl | gi\|119874718 | 69.3 | 85.96 | 9.00E-42 | 174 |
| gi\|9230677\|gl | gi\|213421212 | 97.59 | 98.8 | 1.00E-39 | 167 |
| gi\|9230677\|gl | gi\|238786276 | 47.93 | 65.68 | 1.00E-38 | 164 |
| gi\|9230677\|gl | gi\|238761968 | 50.31 | 66.46 | 2.00E-38 | 164 |
| gi\|9230677\|gl | gi\|213027082 | 100 | 100 | 1.00E-35 | 154 |
| gi\|9230677\|gl | gi\|51246554\|r | 32.27 | 51.77 | 1.00E-33 | 147 |
| gi\|9230677\|gl | gi\|229904894 | 77.11 | 84.34 | 2.00E-29 | 134 |
| gi\|9230677\|gl | gi\|32250733\|( | 77.5 | 85 | 5.00E-28 | 129 |
| gi\|9230677\|gl | gi\|293386382 | 59.6 | 72.73 | 2.00E-25 | 120 |

2. Complete Protein Sequence Of Kgp

NCBI Reference Sequence: YP_001929844.1

## lysine-specific cysteine proteinase Kgp [Porphyromonas gingivalis ATCC 33277]

```
>gi|188995592|ref|YP_001929844.1| lysine-specific cysteine proteinase Kgp
[Porphyromonas gingivalis ATCC 33277]
MRKLLLLIAASLLGVGLYAQSAKIKLDAPTTRTTCTNNSFKQFDASFSFNEVELTKVETKGGTFASVSIP
GAFPTGEVGSPEVPAVRKLIAVPVGATPVVRVKSFTEQVYSLNQYGSEKLMPHQPSMSKSDDPEKVPFVY
NAAAYARKGFVGQELTQVEMLGTMRGVRIAALTINPVQYDVVANQLKVRNNIEIEVSFQGADEVATQRLY
DASFSPYFETAYKQLFNRDVYTDHGDLYNTPVRMLVVAGAKFKEALKPWLTWKAQKGFYLDVHYTDEAEV
GTTNASIKAFIHKKYNDGLAASAAPVFLALVGDTDVISGEKGKKTKKVTDLYYSAVDGDYFPEMYTFRMS
ASSPEELTNIIDKVLMYEKATMPDKSYLEKALLIAGADSYWNPKIGQQTIKYAVQYYYNQDHGYTDVYSY
PKAPYTGCYSHLNTGVGFANYTAHGSETSWADPSLTATQVKALTNKDKYFLAIGNCCVTAQFDYPQPCFG
EVMTRVKEKGAYAYIGSSPNSYWGEDYYWSVGANAVFGVQPTFEGTSMGSYDATFLEDSYNTVNSIMWAG
NLAATHAGNIGNITHIGAHYYWEAYHVLGDGSVMPYRAMPKTNTYTLPASLPQNQASYSIQASAGSYVAI
SKDGVLYGTGVANASGVATVNMTKQITENGNYDVVITRSNYLPVIKQIQAGEPSPYQPVSNLTATTQGQK
VTLKWDAPSAKKAEASREVKRIGDGLFVTIEPANDVRANEAKVVLAADNVWGDNTGYQFLLDADHNTFGS
VIPATGPLFTGTASSNLYSANFEYLIPANADPVVTTQNIIVTGQGEVVIPGGVYDYCITNPEPASGKMWI
AGDGGNQPARYDDFTFEAGKKYTFTMRRAGMGDGTDMEVEDDSPASYTYTVYRDGTKIQEGLTATTFEED
GVAAGNHEYCVEVKYTAGVSPKVCKDVTVEGSNEFAPVQNLTGSAVGQKVTLKWDAPNGTPNPNPNPNPG
TTTLSESFENGIPASWKTIDADGDGHGWKPGNAPGIAGYNSNGCVYSESFGLGGIGVLTPDNYLITPALD
LPNGGKLTFWVCAQDANYASEHYAVYASSTGNDASNFTNALLEETITAKGVRSPEAIRGRIQGTWRQKTV
DLPAGTKYVAFRHFQSTDMFYIDLDEVEIKANGKRADFTETFESSTHGEAPAEWTTIDADGDGQDWLCLS
SGQLDWLTAHGGTNVVASFSWNGMALNPDNYLISKDVTGATKVKYYYAVNDGFPGDHYAVMISKTGTNAG
DFTVVFEETPNGINKGGARFGLSTEANGAKPQSVWIERTVDLPAGTKYVAFRHYNCSDLNYILLDDIQFT
MGGSPTPTDYTYTVYRDGTKIKEGLTETTFEEDGVATGNHEYCVEVKYTAGVSPKVCVNVTINPTQFNPV
KNLKAQPDGGDVVLKWEAPSGKRGELLNEDFEGDAIPTGWTALDADGDGNNWDITLNEFTRGERHVLSPL
RASNVAISYSSLLQGQEYLPLTPNNFLITPKVEGAKKITYKVGSPGLPQWSHDHYALCISKSGTAAADFE
VIFEETMTYTQGGANLTREKDLPAGTKYVAFRHYNCTDVLGIMIDDVVITGEGEGPSYTYTVYRDGTKIQ
EGLTETTYRDAGMSAQSHEYCVEVKYAAGVSPKVCVDYIPDGVADVTAQKPYTLTVVGKTITVTCQGEAM
IYDMNGRRLAAGRNTVVYTAQGGYYAVMVVVDGKSYVEKLAIK
```

3. Complete Sequence of RgpA

NCBI Reference Sequence: YP_001930085.1

## arginine-specific cysteine proteinase RgpA [Porphyromonas gingivalis ATCC 33277]

```
>gi|188995833|ref|YP_001930085.1| arginine-specific cysteine proteinase RgpA
[Porphyromonas gingivalis ATCC 33277]
MNKFVSIALCSSLLGGMAFAQQTELGRNPNVRLLESTQQSVTKVQFRMDNLKFTEVQTPKGMAQVPTYTE
GVNLSEKGMPTLPILSRSLAVSDTREMKVEVVSSKFIEKKNVLIAPSKGMIMRNEDPKKIPYVYGKSYSQ
NKFFPGEIATLDDPFILRDVRGQVVNFAPLQYNPVTKTLRIYTEITVAVSETSEQGKNILNKKGTFAGFE
DTYKRMFMNYEPGRYTPVEEKQNGRMIVIVAKKYEGDIKDFVDWKNQRGLRTEVKVAEDIASPVTANAIQ
QFVKQEYEKEGNDLTYVLLVGDHKDIPAKITPGIKSDQVYGQIVGNDHYNEVFIGRFSCESKEDLKTQID
RTIHYERNITTEDKWLGQALCIASAEGGPSADNGESDIQHENVIANLLTQYGYTKIIKCYDPGVTPKNII
DAFNGGISLVNYTGHGSETAWGTSHFGTTHVKQLTNSNQLPFIFDVACVNGDFLFSMPCFAEALMRAQKD
GKPTGTVAIIASTINQSWASPMRGQDEMNEILCEKHPNNIKRTFGGVTMNGMFAMVEKYKKDGEKMLDTW
TVFGDPSLLVRTLVPTKMQVTAPAQINLTDASVNVSCDYNGAIATISANGKMFGSAVVENGTATINLTGL
TNESTLTLTVVGYNKETVIKTINTNGEPNPYQPVSNLTATTQGQKVTLKWDAPSTKTNATTNTARSVDGI
RELVLLSVSDAPELLRSGQAEIVLEAHDVWNDGSGYQILLDADHDQYGQVIPSDTHTLWPNCSVPANLFA
PFEYTVPENADPSCSPTNMIMDGTASVNIPAGTYDFAIAAPQANAKIWIAGQGPTKEDDYVFEAGKKYHF
LMKKMGSGDGTELTISEGGGSDYTYTVYRDGTKIKEGLTATTFEEDGVATGNHEYCVEVKYTAGVSPKVC
KDVTVEGSNEFAPVQNLTGSAVGQKVTLKWDAPNGTPNPNPNPNPNPNPGTTTLSESFENGIPASWKTID
ADGDGHGWKPGNAPGIAGYNSNGCVYSESFGLGGIGVLTPDNYLITPALDLPNGGKLTFWVCAQDANYAS
EHYAVYASSTGNDASNFTNALLEETITAKGVRSPEAIRGRIQGTWRQKTVDLPAGTKYVAFRHFQSTDMF
YIDLDEVEIKANGKRADFTETFESSTHGEAPAEWTTIDADGDGQGWLCLSSGQLDWLTAHGGTNVVASFS
WNGMALNPDNYLISKDVTGATKVKYYYAVNDGFPGDHYAVMISKTGTNAGDFTVVFEETPNGINKGGARF
GLSTEANGAKPQSVWIERTVDLPAGTKYVAFRHYNCSDLNYILLDDIQFTMGGSPTPTDYTYTVYRDGTK
IKEGLTETTFEEDGVATGNHEYCVEVKYTAGVSPKECVNVTINPTQFNPVKNLKAQPDGGDVVLKWEAPS
AKKTEGSREVKRIGDGLFVTIEPANDVRANEAKVVLAADNVWGDNTGYQFLLDADHNTFGSVIPATGPLF
TGTASSNLYSANFEYLIPANADPVVTTQNIIVTGQGEVVIPGGVYDYCITNPEPASGKMWIAGDGGNQPA
RYDDFTFEAGKKYTFTMRRAGMGDGTDMEVEDDSPASYTYTVYRDGTKIKEGLTETTYRDAGMSAQSHEY
CVEVKYAAGVSPKVCVDYIPDGVADVTAQKPYTLTVVGKTITVTCQGEAMIYDMNGRRLAAGRNTVVYTA
QGGYYAVMVVVDGKSYVEKLAVK
```

4. Complete sequence of RgpB

NCBI Reference Sequence: YP_001929582.1

## arginine-specific cysteine proteinase RgpB [Porphyromonas gingivalis ATCC 33277]

```
>gi|188995330|ref|YP_001929582.1| arginine-specific cysteine proteinase RgpB
[Porphyromonas gingivalis ATCC 33277]
MKKNFSRIVSIVAFSSLLGGMAFAQPAERGRNPQVRLLSAEQSMSKVQFRMDNLQFTDVQTSKGVAQVPT
FTEGVNISEKGTPILPILSRSLAVSETRAMKVEVVSSKFIEKKDVLIAPSKGVISRAENPDQIPYVYGQS
YNEDKFFPGEIATLSDPFILRDVRGQVVNFAPLQYNPVTKTLRIYTEIVVAVSETAEAGQNTISLVKNST
FTGFEDIYKSVFMNYEATRYTPVEEKENGRMIVIVAKKYEGDIKDFVDWKNQRGLRTEVKVAEDIASPVT
ANAIQQFVKQEYEKEGNDLTYVLLVGDHKDIPAKITPGIKSDQVYGQIVGNDHYNEVFIGRFSCESKEDL
KTQIDRTIHYERNITTEDKWLGQALCIASAEGGPSADNGESDIQHENVIANLLTQYGYTKIIKCYDPGVT
PKNIIDAFNGGISLVNYTGHGSETAWGTSHFGTTHVKQLTNSNQLPFIFDVACVNGDFLFSMPCFAEALM
RAQKDGKPTGTVAIIASTINQSWASPMRGQDEMNEILCEKHPNNIKRTFGGVTMNGMFAMVEKYKKDGEK
MLDTWTVFGDPSLLVRTLVPTEMQVTAPANISASAQTFEVACDYNGAIATLSDDGDMVGTAIVKDGKAII
KLNESIADETNLTLTVVGYNKVTVIKDVKVEGTSIADVANDKPYTVAVSGKTITVESPAAGLTIFDMNGR
RVATAKNRMVFEAQNGVYAVRIATEGKTYTEKVIVK
```